\documentclass[]{aa}

\usepackage{graphics}

\usepackage{natbib}
\bibpunct{(}{)}{;}{a}{}{,}

\def\msun{{\rm\,M_\odot}}
\def\be{\begin{equation}}
\def\ee{\end{equation}}
\def\hompc{\,h\,{\rm Mpc}^{-1}}

\begin{document}

\title{Cosmological structure formation in a homogeneous dark energy background}

\titlerunning{Cosmological structure formation with dark energy}

\author{Will J. Percival\inst{1,2}}

\institute{
$^1$ Institute for Astronomy, University of Edinburgh, Blackford
Hill, Edinburgh EH9 3HJ\\
$^2$ Institute of Cosmology and Gravitation, University of Portsmouth,
Portsmouth, P01 2EG\\ 
\email{wjp@roe.ac.uk}}

\date{\today}

\abstract{There is now strong evidence that the current energy density
of the Universe is dominated by dark energy with an equation of state
$w<-1/3$, which is causing accelerated expansion. The build-up of
structure within such Universes is subject to significant ongoing
study, particularly through the spherical collapse model. This paper
aims to review and consolidate progress for cosmologies in which the
dark energy component remains homogeneous on the scales of the
structures being modelled. The equations presented are designed to
allow for dark energy with a general time-varying equation of state
$w(a)$. In addition to reviewing previous work, a number of new
results are introduced: A new fitting formula for the linear growth
factor in constant $w$ cosmologies is given. A new formalism for
determining the critical density for collapse is presented based on
the initial curvature of the perturbation. The commonly used
approximation to the critical density for collapse based on the linear
growth factor is discussed for a general dark energy equation of
state. Virialisation within such cosmologies is also considered, and
the standard assumption that energy is conserved between turn-around
and virialisation is questioned and limiting possiblities are
presented.  \keywords{Cosmology: theory -- large-scale structure of
Universe} }

\maketitle

\section{Introduction}

Analysis of the distance-redshift relation using high redshift Type Ia
supernovae has led to the discovery that the expansion of the Universe
is currently accelerating \citep{riess98,perlmutter99}. This suggests
that the dominant contribution to the present-day energy budget is a
component with equation of state $w<-1/3$, called
``dark-energy''. Combining measurements of CMB fluctuations
\citep{debernardis00,lee01,halverson02,pearson03,scott03,hinshaw03}
with measurements of the clustering of present day galaxies
\citep{percival01,tegmark04a,cole05} has confirmed this requirement
for dark energy
(e.g. \citealt{efstathiou02,percival02,spergel03,tegmark04b}). These
studies favour a flat Universe with $\Omega_M\simeq0.3$, with the
remaining contribution made up of dark energy.

The nature of the dark energy is the source of much debate. Perhaps
the most straightforward candidate is a positive cosmological constant
$\Lambda$ with equation of state parameter $w=-1$. This simple picture
forms a special case in a broader class of models where the dark
energy is the manifestation of a scalar field slowly rolling down its
potential. In the limit of a completely flat potential, these models
lead to $w=-1$ \citep{wetterich88,peebles88,ratra88}. A subclass of
quintessence models with constant $-1<w<-1/3$ was proposed by
\citet{caldwell98a,caldwell98b}. Many other forms have been proposed
for the shape of this potential, leading to an equation of state
parameter that is dependent on the scale factor (see
\citealt{peebles03} for a review).

Given the plethora of models, many parameterisations have been
introduced in order to observationally constrain $w(a)$ (see
\citealt{johri04} for a review). However, there is obviously a
limitation imposed by any parameterisation \citep{bassett04}
and, in the present paper, we have tried to be as general as possible
in the equations presented and discussed. In order to demonstrate the
effects of the dark energy equation of state in numerical examples, we
pay particular attention to the parameterisation of \citet{jassal04}
\be
  w(a)=w_0+w_1a(1-a).
    \label{eq:wparam}
\ee
The present day value of $w|_{a=1}=w_0$, and its derivative
$dw(a)/da|_{a=1}=-w_1$ will serve to demonstrate the possible
cosmological signatures of a wider class of models. We also consider
models that assume $w(a)=w$ is constant.  Qualitatively, many of the
effects of general $w(a)$ models can be predicted by interpolating
between models with constant $w$. In fact, for many models, by the
lookback time by which $w(a)$ has evolved significantly from its
present day value, the cosmological significance of $w(a)$ is greatly
reduced (although see, for example, \citealt{maor02}). Obviously, this
also signifies that an evolving $w(a)$ is harder to
observationally constrain than constant $w$ \citep{kujat02}. If the
dark energy equation of state only varies slowly with time, then
observational predictions are well approximated by treating $w(a)=w$
as a constant \citep{wang00}, with
\be
  w \simeq \frac{\int\,da\, \Omega_X(a)w(a)}{\int\,da\, \Omega_X(a)},
\ee
where $\Omega_X$ is the dark energy density relative to the critical
density (see also \citealt{dave02}).

The equation of state for the dark energy does not uniquely define the
behaviour of this component. The formation of structure is also
dependent on the sound speed of the dark energy which limits its
clustering properties. In the original formalism for quintessence
\citep{caldwell98a,caldwell98b}, the dark energy component has a high
sound speed which means that it can cluster on the largest scales, but
does not cluster on the scales of galaxy clusters and below.

Consequently, the dark energy only affects the matter power
spectrum \citep{ma99}, and the CMB anisotropies \citep{caldwell98b} on
very large scales. For the special case of a cosmological constant,
$w=-1$, the clustering of the dark energy is not an issue as the
energy density in perturbations always remains at the background
level. Obviously, for $w\ne-1$, the clustering properties of the dark
energy strongly affect the build-up of structure in the Universe. In
particular for spherical perturbations, the linear growth rate,
critical overdensity for collapse, and details of subsequent behaviour
and virialisation are all dependent on this property. In this paper we
follow the majority of current literature and only consider a
non-clustering dark energy component. However, we do note that models
in which the dark energy clusters on small scales are being discussed
with increasing frequency
\citep{hu04a,nunes04,hannestad05,manera05,maor05}.

In general, for the variables used in this paper, if no dependence is
quoted for a given quantity (e.g. $\Omega_M$), it should be assumed to
be calculated at present day. If instead explicit dependence is given
(e.g. $\Omega_M(a)$), the quantity is assumed to vary with epoch. The
exception to this rule is $w(a)$, where if no dependence on $a$ is
given, we additionally assume that $w(a)=w$ is constant in time.

We start by reviewing the behaviour of different cosmological models,
and the critical parameters that separate models with different
properties (Section~\ref{sec:dynamics}) as this can be related to the
behaviour of spherical perturbations. We then calculate the linear
growth factor (Section~\ref{sec:linear}), the critical overdensity for
collapse at present day (Section~\ref{sec:dc}), and as a function of
time (Section~\ref{sec:dc_evol}). This critical overdensity is then
used to determine the mass function (Section~\ref{sec:mass_func}) and
the rate of structure growth (Section~\ref{sec:rate_grow}). Finally we
consider the virialisation of perturbations, and the difficulties
associated with these calculations in dark energy cosmologies
(Section~\ref{sec:virial}).

\section{dynamics of cosmological models}  \label{sec:dynamics}

It is assumed that the dark energy has an equation of state relating
its pressure $p_X$ and density $\rho_X$ given by
$p_X=w(a)\rho_X$. For general $w(a)$, the dynamical expansion of the
Universe is specified by the Friedmann equation
\be
  E^2(a) = \frac{H^2(a)}{H_0^2}
         = \Omega_Ma^{-3}+\Omega_Ka^{-2}+\Omega_Xa^{f(a)},
  \label{eq:friedmann_eq1}
\ee
where $\Omega_K\equiv(1-\Omega_M-\Omega_X)$ is the curvature constant,
$H(a)\equiv\dot{a}/a$ is the Hubble parameter with present day value
$H_0$. $f(a)$ is calculated by solving the conservation of energy
equation for the dark energy $d(\rho_Xa^3)/da = -3p_Xa^2$ \citep{caldwell98b}, 
giving $\rho_X\propto a^{f(a)}$, where
\be
  f(a) = \frac{-3}{\ln a}\int_0^{\ln a}[1+w(a')]d\ln a'.
   \label{eq:rho_X}
\ee
For constant $w$, $f(a)=-3(1+w)$. For the parameterisation
$w(a)=w_0+w_1a(1-a)$ of \citet{jassal04}, 
\be
  f(a)=-3(1+w_0)+\frac{3w_1}{2\ln a}(1-a)^2.
  \label{eq:faw_jassal}
\ee

The evolution of the matter density $\Omega_M(a)$ and dark energy
density $\Omega_X(a)$ are given by
\be
  \Omega_M(a)=\frac{\Omega_M a^{-3}}{E^2(a)},\,\,\,
  \Omega_X(a)=\frac{\Omega_X a^{f(a)}}{E^2(a)}.
  \label{eq:omega_evol}
\ee 

It is immediately apparent that the equation of state of the dark
energy has a strong affect on the behaviour of these equations, even
for models where $w(a)$ is constant. For example, for $w=-1/3$, the
dark energy terms in Eq.~\ref{eq:friedmann_eq1} cancel, leaving a
cosmological model with an expansion time history that behaves exactly
as an open Universe with matter density $\Omega_M$, although the
space-time geometry differs (e.g. review by \citealt{peebles03}).  For
$w<-1/3$, at late times the dark energy will dominate the dynamics of
the expansion, with the epoch of transition from matter to dark energy
domination dependent on $w$. Decreasing the equation of state $w$ from
$-1/3$ smoothly interpolates between the open Universe model,
cosmological constant ($\Lambda$ model) and extrapolates beyond.

The asymptotic behaviour (either forwards or backwards in time) of
different cosmological models is delineated by models with so-called
critical parameters. These separate cosmologies that predict future
recollapse, expansion forever and models that do not start at a
big-bang, but instead ``loiter'' at early times, asymptotically
tending towards a stable solution such as Einstein's static model for
$\Lambda$ cosmologies. For these critical models, the zero points of
$(da/dt)^2=E^2(a)a^2=0$ should occur at turning points of this
equation. For constant $w$, differentiating
Eq.~\ref{eq:friedmann_eq1}, we see that the critical cosmologies
should have $E^2(a)a^2=0$ when \citep{chiba05}
\be 
  a_{\rm turn} = \left[\frac{\Omega_M}{-(1+3w)\Omega_X}
    \right]^{\frac{-1}{3w}}.
    \label{eq:a_turn}
\ee
Substituting this value of the scale factor back into
Eq.~\ref{eq:friedmann_eq1} gives an equation for the critical
parameters. 

For $w=-1$, this reduces to the well known cubic equation
\citep{glanfield66} 
\be
  (\Omega_M+\Omega_\Lambda-1)^3 = \frac{27}{4}\Omega_M^2\Omega_\Lambda
  \label{eq:omega_critical}
\ee
that can be solved analytically \citep{felten86}. A detailed
discussion of the behaviour of $\Lambda$ models is given by
\citet{moles91}.

\begin{figure*}
  \resizebox{5.8cm}{!}{\includegraphics{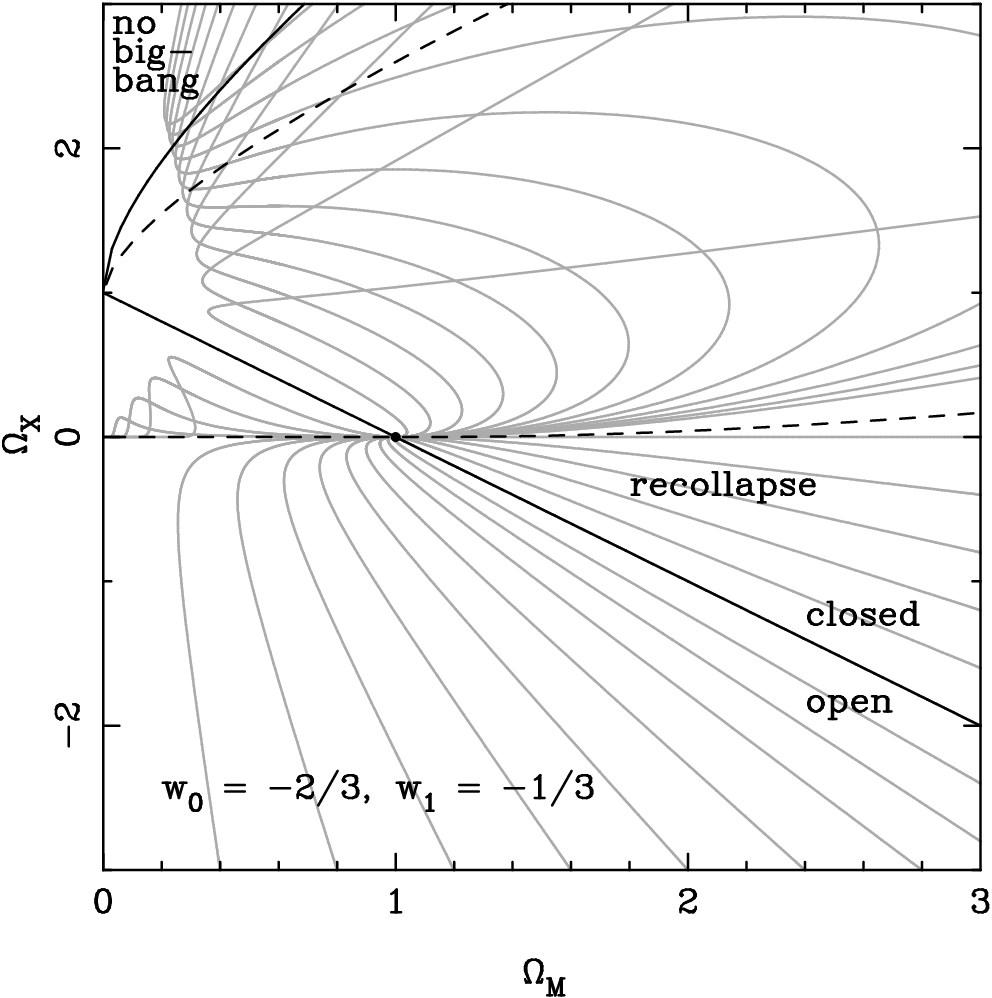}}
  \resizebox{5.8cm}{!}{\includegraphics{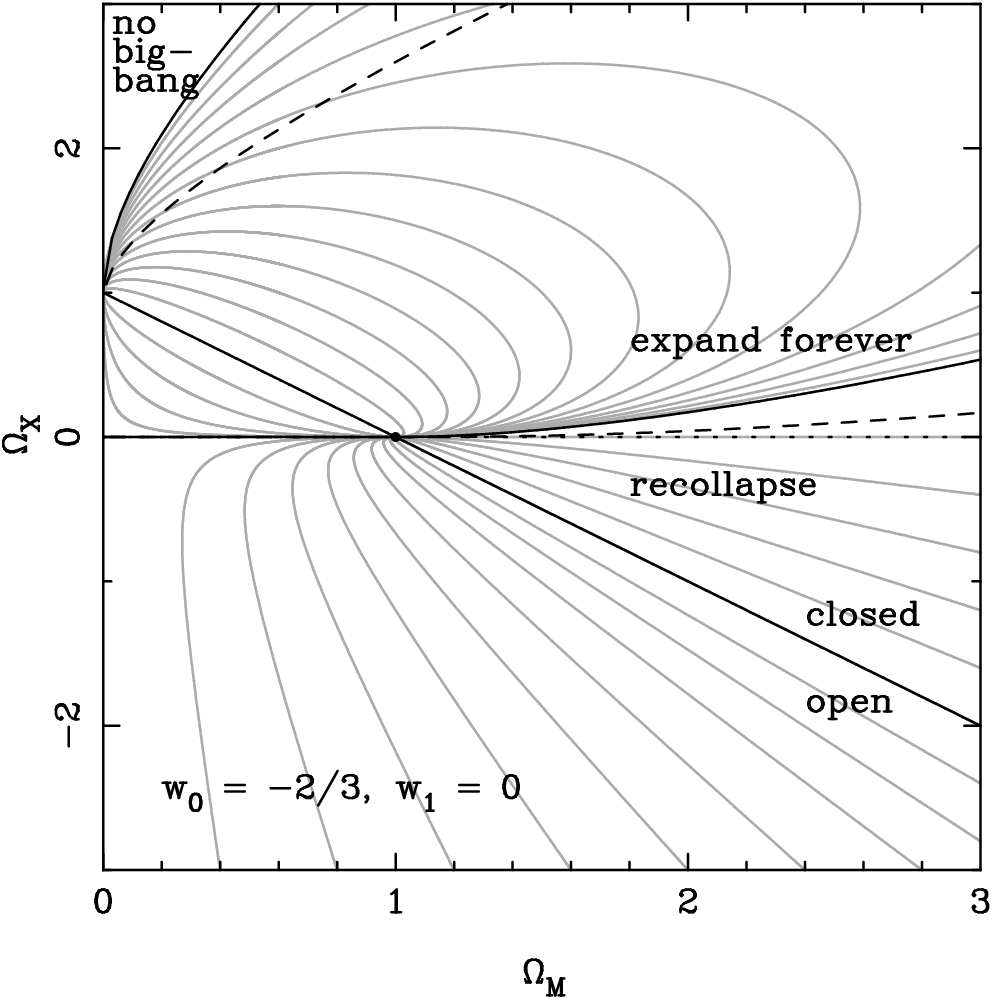}}
  \resizebox{5.8cm}{!}{\includegraphics{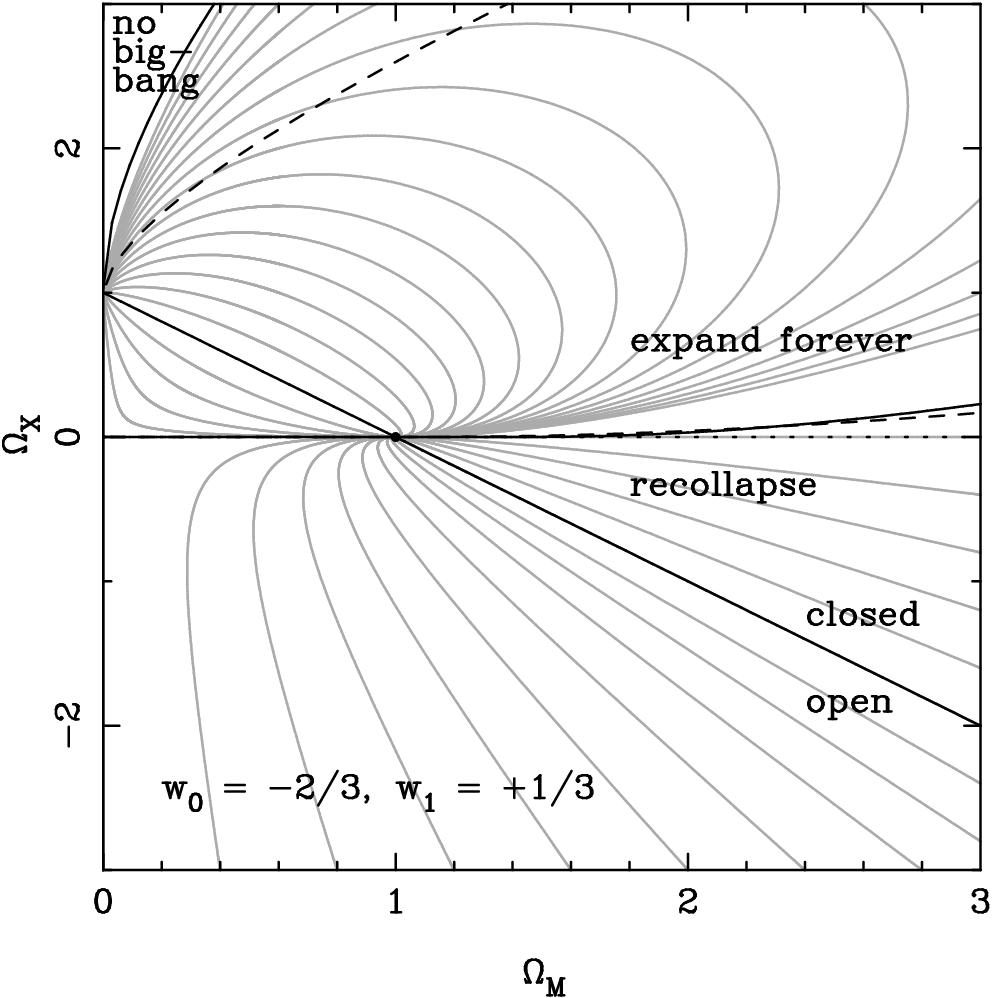}}\\
  \resizebox{5.8cm}{!}{\includegraphics{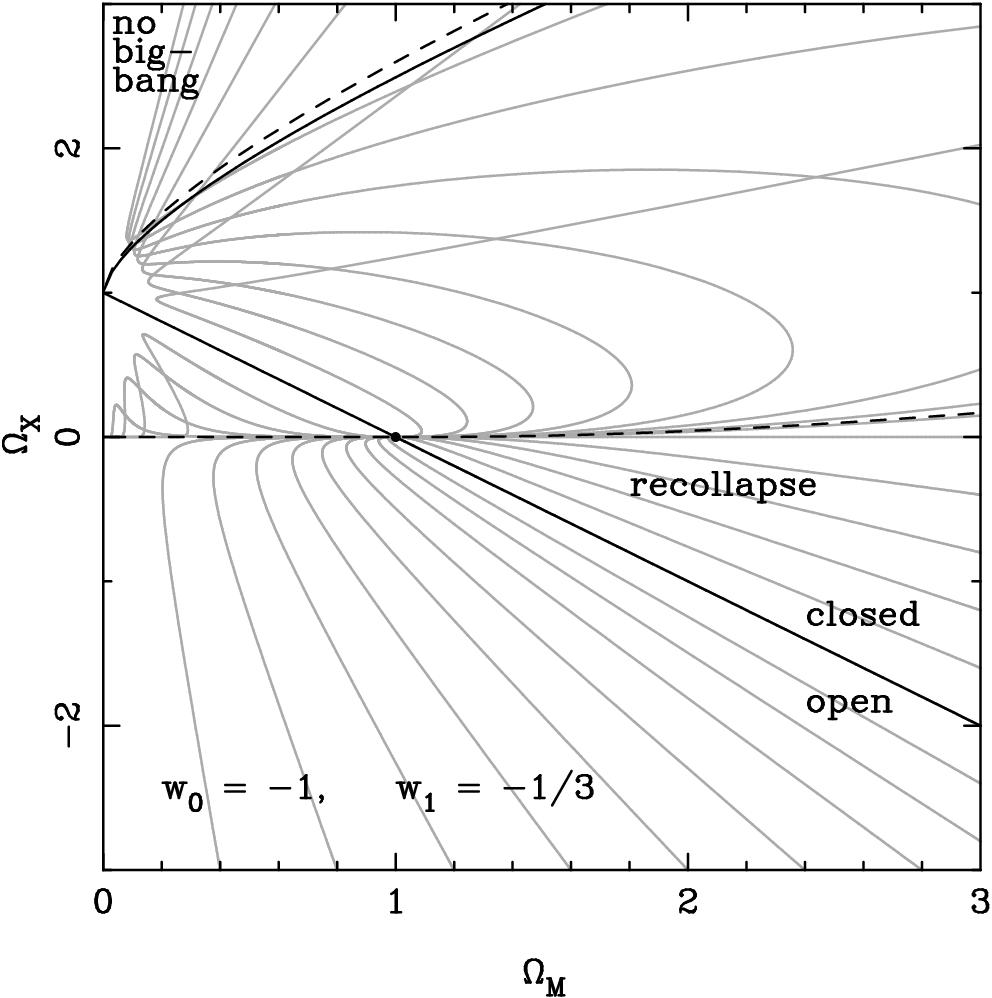}}
  \resizebox{5.8cm}{!}{\includegraphics{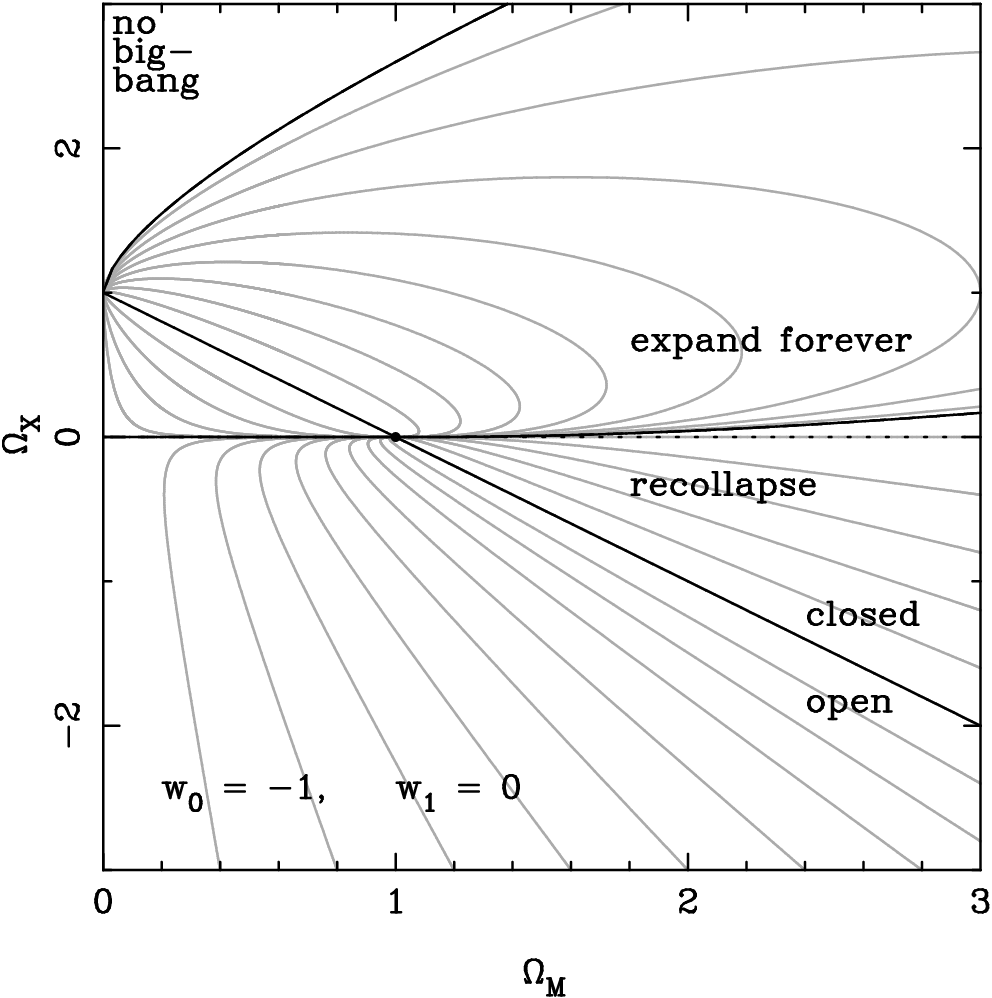}}
  \resizebox{5.8cm}{!}{\includegraphics{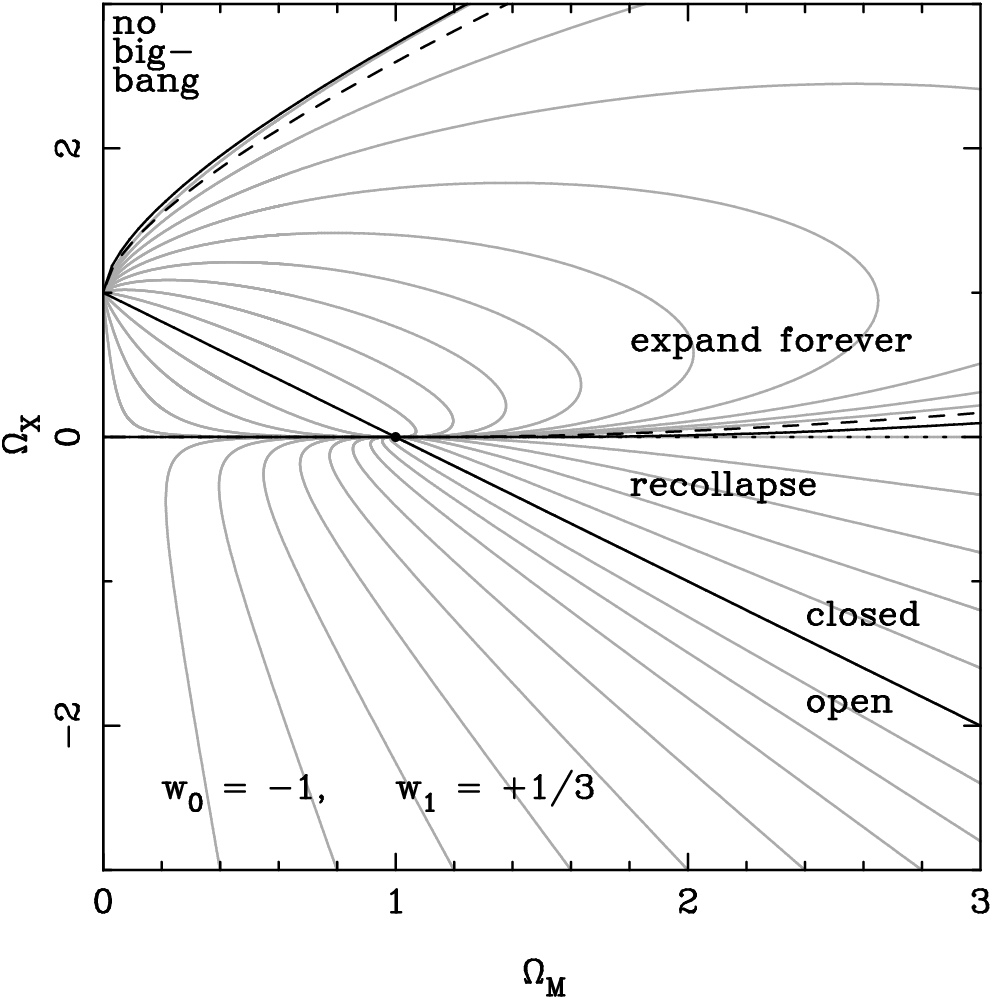}}\\
  \resizebox{5.8cm}{!}{\includegraphics{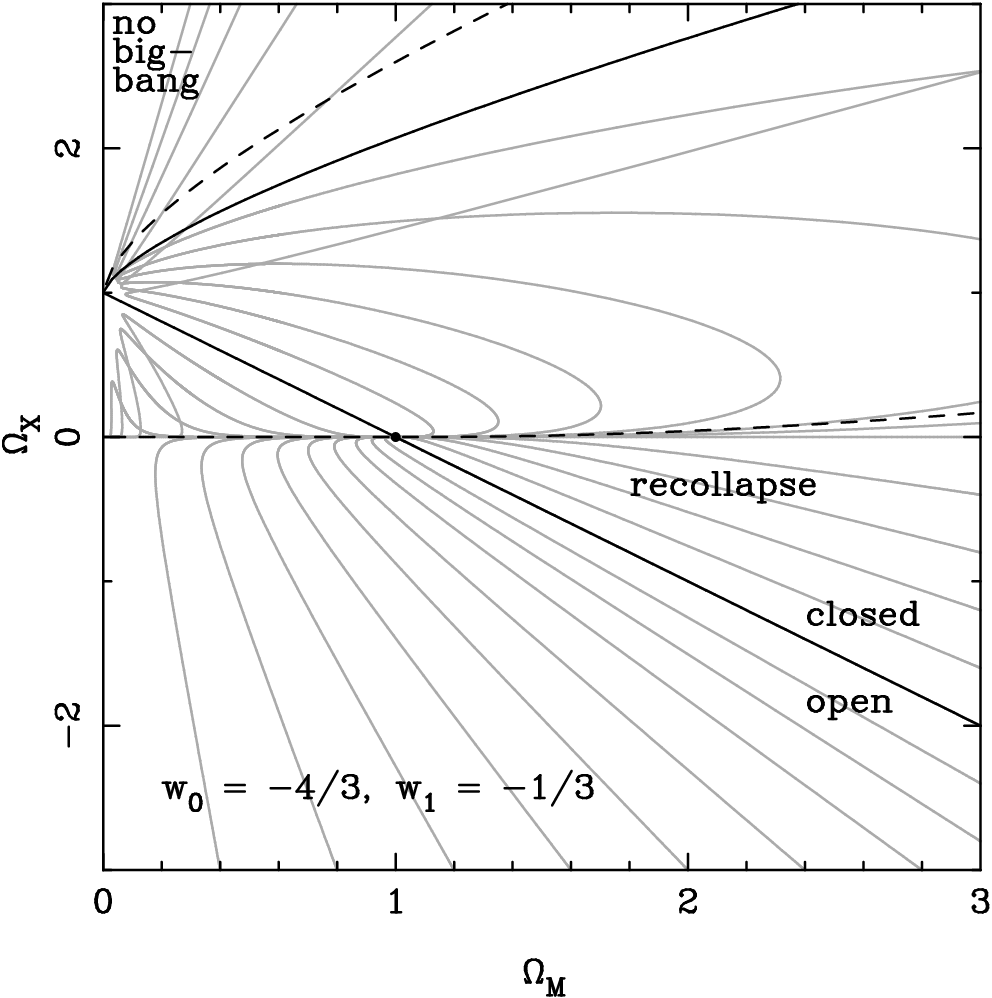}}
  \resizebox{5.8cm}{!}{\includegraphics{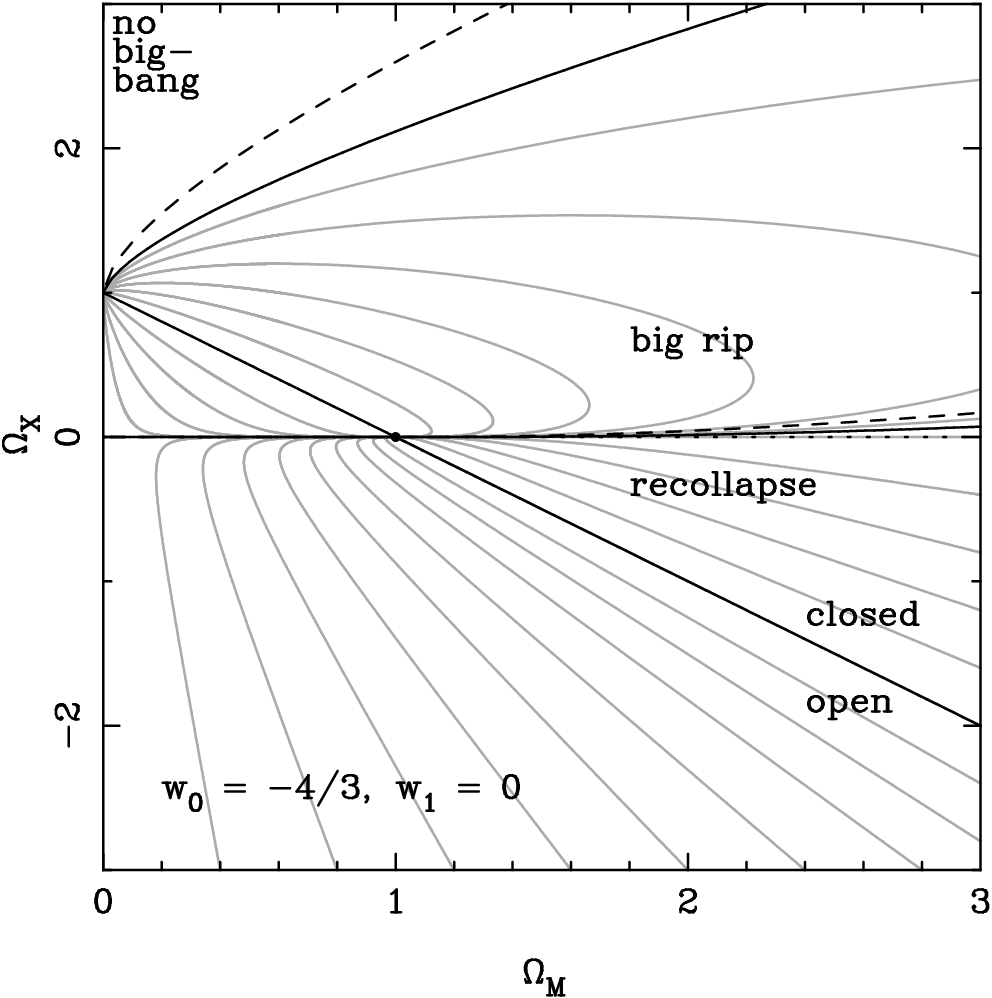}}
  \resizebox{5.8cm}{!}{\includegraphics{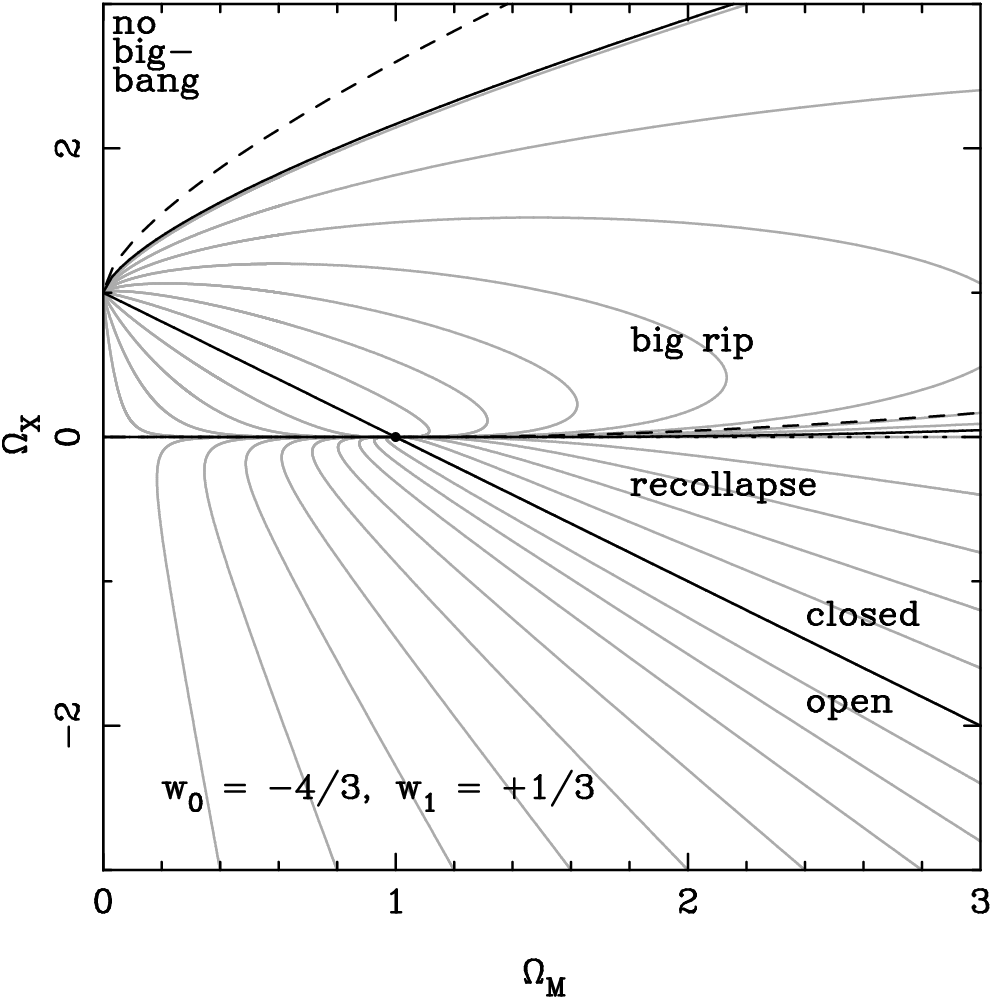}}\\
  \caption{Plots showing the evolution of the matter and vacuum energy
  densities for a selection of cosmologies with different dark energy
  equation of state (grey lines). For cosmologies with $\Omega_X>0$,
  $a=1$ was fixed when the matter density and vacuum density were
  equal. The critical models that border the different types of
  evolution are shown by the black lines. Dashed lines in all plots
  show the critical models for $\Lambda$ cosmologies. The dotted line
  for $w_1\ge0$ shows $\Omega_X=0$, to emphasis that recollapse can
  occur if $\Omega_X>0$ provided that $\Omega_M>>\Omega_X$.
  \label{fig:critical_cosmology_models}}
\end{figure*}

For general $w(a)$, it is straightforward to numerically determine the
critical parameters from Eq.~\ref{eq:friedmann_eq1} and its
derivative. The different regions in ($\Omega_M$,$\Omega_X$) space are
shown in Fig.~\ref{fig:critical_cosmology_models} for $w_0=-4/3,-1,
-2/3$ and $w_1=-1/3,0,1/3$ using the parameterisation of
Eq.~\ref{eq:wparam}. Critical values of $\Omega_M$ and $\Omega_X$ at
$a=1$ are shown by the solid black lines.

For comparison, in Fig.~\ref{fig:critical_cosmology_models} we also
plot various cosmological tracks (grey lines). For general $w(a)$
models, the cosmology is not uniquely specified by $\Omega_M(a)$ and
$\Omega_X(a)$, as we also need to specify $a$. The cosmological tracks
in Fig.~\ref{fig:critical_cosmology_models} for $\Omega_X>0$ have
$a=1$ when $\Omega_M=\Omega_X$. For general models, the tracks can
therefore cross, and can move into regions of
($\Omega_M(a)$,$\Omega_X(a)$) space that would have been excluded at
$a=1$. For the panels where $w_1\ne0$, the cosmological tracks should
simply be though of as examples of possible tracks that go through
particular $\Omega_M(a)$ and $\Omega_X(a)$.

For $w_1=0$, $w_0<-1$ models favour a larger region of parameter space
for which there is no big-bang, while $w_0>-1$ leads to recollapse for
models in a larger region in ($\Omega_M$, $\Omega_X$) space,
reflecting the fact that a smaller amount of matter is required to
collapse the Universe before the dark energy dominates. For $w_0>-1$
cosmologies, a future singularity is predicted where the scale factor
and Hubble parameter diverge, leading to a ``big-rip'' rather than
expansion forever \citep{caldwell03a}. For further discussion of the
behaviour of $w_1=0$ cosmologies, see \citet{dearaujo05,chiba05}.

If we allow $w_1\ne0$, the behaviour becomes even more
complicated. From Eq.~\ref{eq:faw_jassal}, for $w_1<0$,
$lim_{a\to\infty}\rho_X=0$, and the dark energy term in
Eq.~\ref{eq:friedmann_eq1} tends to zero. This means that all closed
cosmologies ($\Omega_K<0$) will eventually recollapse as $da/dt=0$ for
some finite $a>1$. Open cosmologies expand forever, with
$lim_{a\to\infty}da/dt=\sqrt{\Omega_K}$. When $\Omega_X>0$, open
models become matter dominated at late times, and can go through a
phase where $\Omega_M(a)$ increases (although obviously $\rho_M$ does
not). For $w_1>0$ and $\Omega_X>0$, the dark energy increases
monotonically with the scale factor, and it is increasingly unlikely
that the matter can cause recollapse before the dark energy takes hold
and accelerates the expansion of the Universe. The region of models
which have $\Omega_X>0$, but that recollapse therefore becomes
smaller.

Solutions that recollapse are of particular importance for the
spherical top-hat collapse model: for $\Lambda$ cosmologies, overdense
regions that recollapse exactly follow the behaviour of these
models. As we will see later on, for general $w(a)$ cosmologies we
cannot link spherical perturbations directly to one of these
cosmological models if the dark energy does not cluster on the scales
of interest.

Given present observational constraints, the region of
$(\Omega_M,\Omega_X)$ space probed in
Fig.~\ref{fig:critical_cosmology_models} is only of academic interest,
and in the remainder of this paper we focus on the subset of models
with $0<\Omega_M<1$ and $0<\Omega_X<1$.

\section{linear growth of fluctuations}  \label{sec:linear}

Considering the behaviour of homogeneous spherical perturbations
provides one of the most simple models for the formation of structure
in the Universe. The ease with which the behaviour can be modelled
follows from Birkhoff's theorem, which states that a spherically
symmetric gravitational field in empty space is static and is always
described by the Schwarzchild metric \citep{birkhoff23}. This gives
that the behaviour of an homogeneous sphere of uniform density can
itself be modelled using the same equations of
Section~\ref{sec:dynamics}. One of the important applications of the
spherical perturbation model is the derivation of the linear growth
rate (pioneered by \citealt{zel58, peebles_lss}, section~10). The
application proceeds as follows: We consider two spheres containing
equal amounts of material, one of background material with radius $a$,
and one of radius $a_p$ with a homogeneous change in
overdensity. Henceforth quantities with a subscript $p$ refer to the
perturbation, while no subscript relates to the background. The
densities within the spheres are related to their radii, with
\be
  \rho_p a_p^3=\rho a^3, \,\,\, \delta\equiv\rho_p/\rho-1,
  \label{eq:arel1}
\ee
giving, to first order in in $\delta$,
\be
  a_p=a(1-\delta/3).
  \label{eq:arel2}
\ee
The cosmological equation for both the spherical perturbation and the
background is
\be
  \frac{1}{a}\frac{d^2a}{dt^2}=-\frac{H_0^2}{2}
    \left[\Omega_Ma^{-3}+[1+3w(a)]\Omega_Xa^{f(a)}\right],
    \label{eq:cosmo}
\ee
where $a$ should be replaced by $a_p$ in the matter density term for
the perturbation. The dark energy density $\rho_X\propto a^{f(a)}$,
is the same for both the perturbation and the background if the dark
energy does not cluster. Because of this, substituting
Eqns~\ref{eq:arel1} \&~\ref{eq:arel2} into this equation gives, to
first order in $\delta$,
\be
  \frac{3}{2}\Omega_MH_0^2a^{-3} \delta = 
  \frac{d^2\delta}{dt^2}+\frac{2}{a}\frac{da}{dt}\frac{d\delta}{dt}.
\ee
Changing variables from $t$ to $a$ gives
\begin{displaymath}
  \frac{3}{2}\Omega_Ma^{-3}\delta = \frac{d^2\delta}{da^2}E^2(a)a^2
    + \frac{d\delta}{da}\left\{2aE^2(a)\right.
\end{displaymath}
\be
  \hspace{1.0cm}
    \left.-\frac{a}{2}\left[\Omega_Ma^{-3}
    +[3w(a)+1]\Omega_Xa^{f(a)}\right]\right\},
  \label{eq:ddda2}
\ee
which can be further simplified to give
\begin{displaymath}
  \frac{3}{2}\Omega_M(a) =   \frac{d^2\ln\delta}{d\ln a^2} 
    + \left(\frac{d\ln\delta}{d\ln a}\right)^2
    + \frac{d\ln\delta}{d\ln a}\left\{1\right.
\end{displaymath}
\be
  \hspace{2cm}
    \left.-\frac{1}{2}\left[\Omega_M(a)+[3w(a)+1]\Omega_X(a)\right]\right\}.
  \label{eq:dlndda2}
\ee
This is the generalisation of Equation B7 in \citet{wang98}
to non-flat cosmologies, and is valid for general $w(a)$. 

\begin{figure}
  \centering
  \resizebox{5.8cm}{!}{\includegraphics{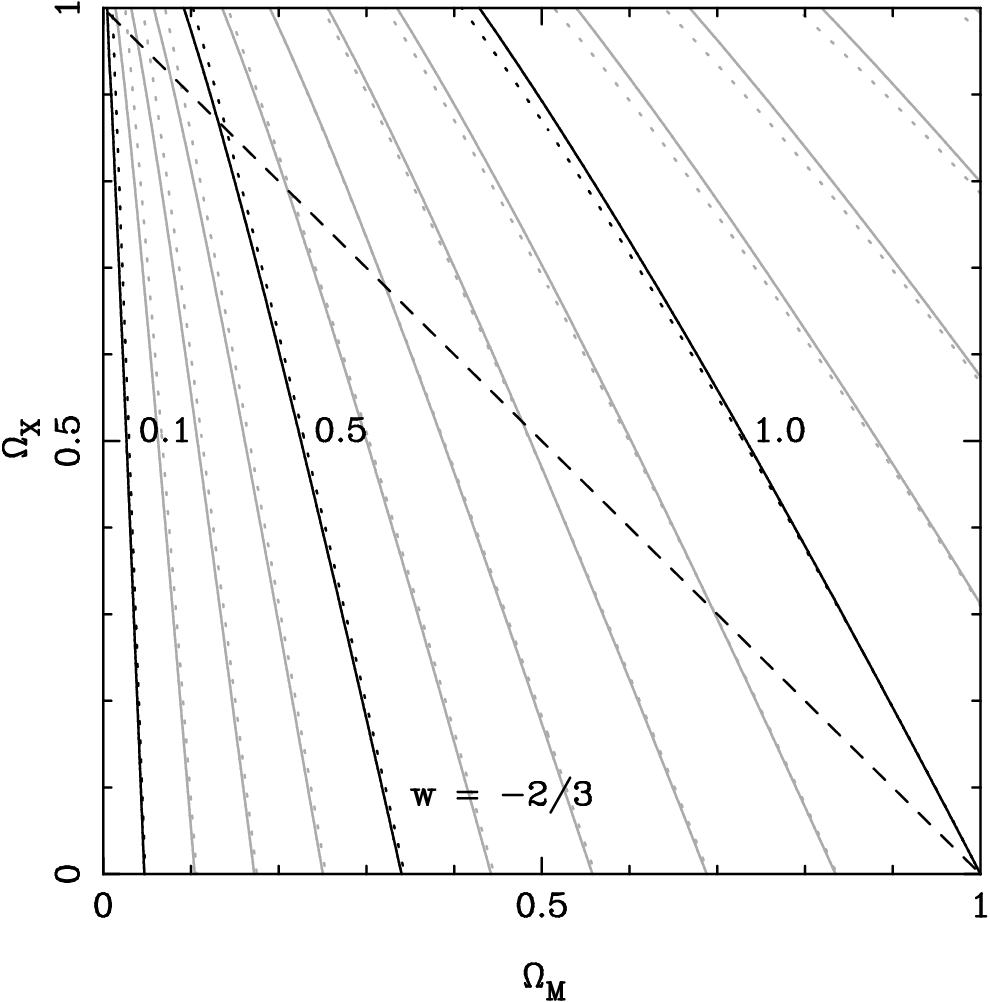}}
  \resizebox{5.8cm}{!}{\includegraphics{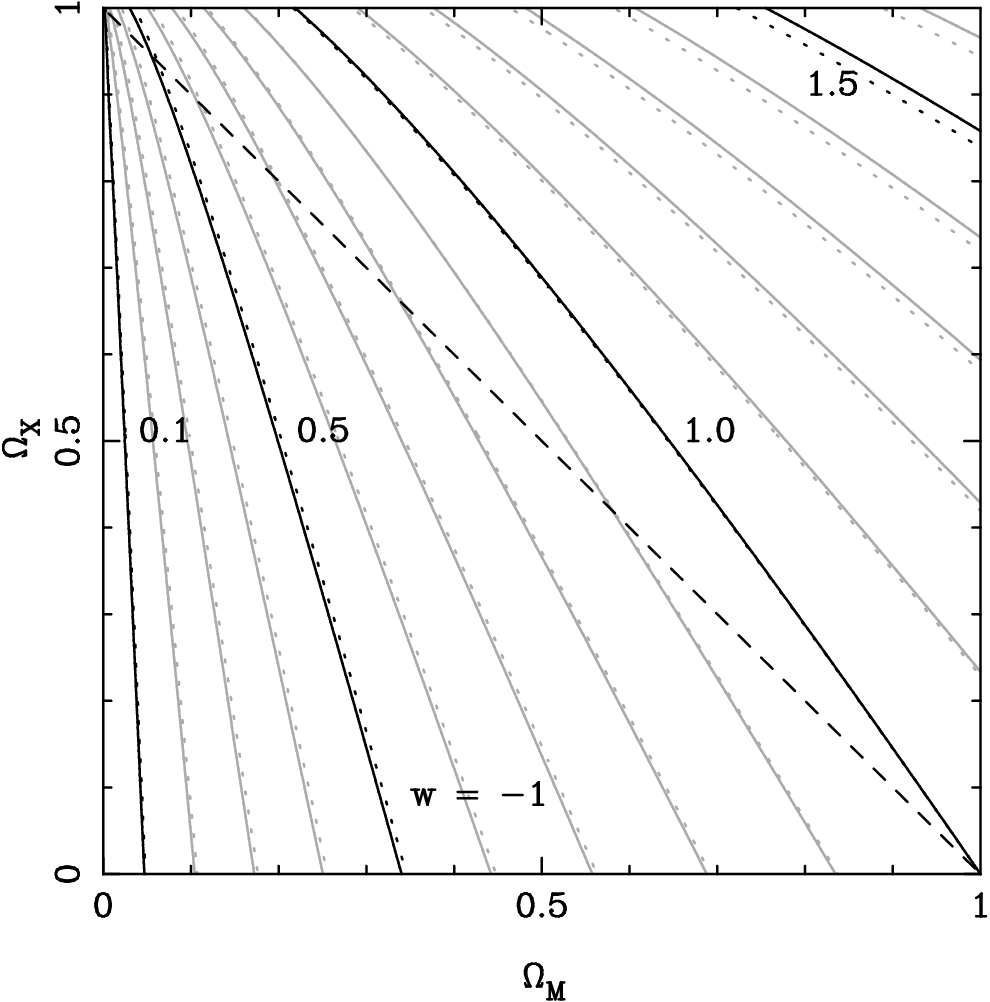}}
  \resizebox{5.8cm}{!}{\includegraphics{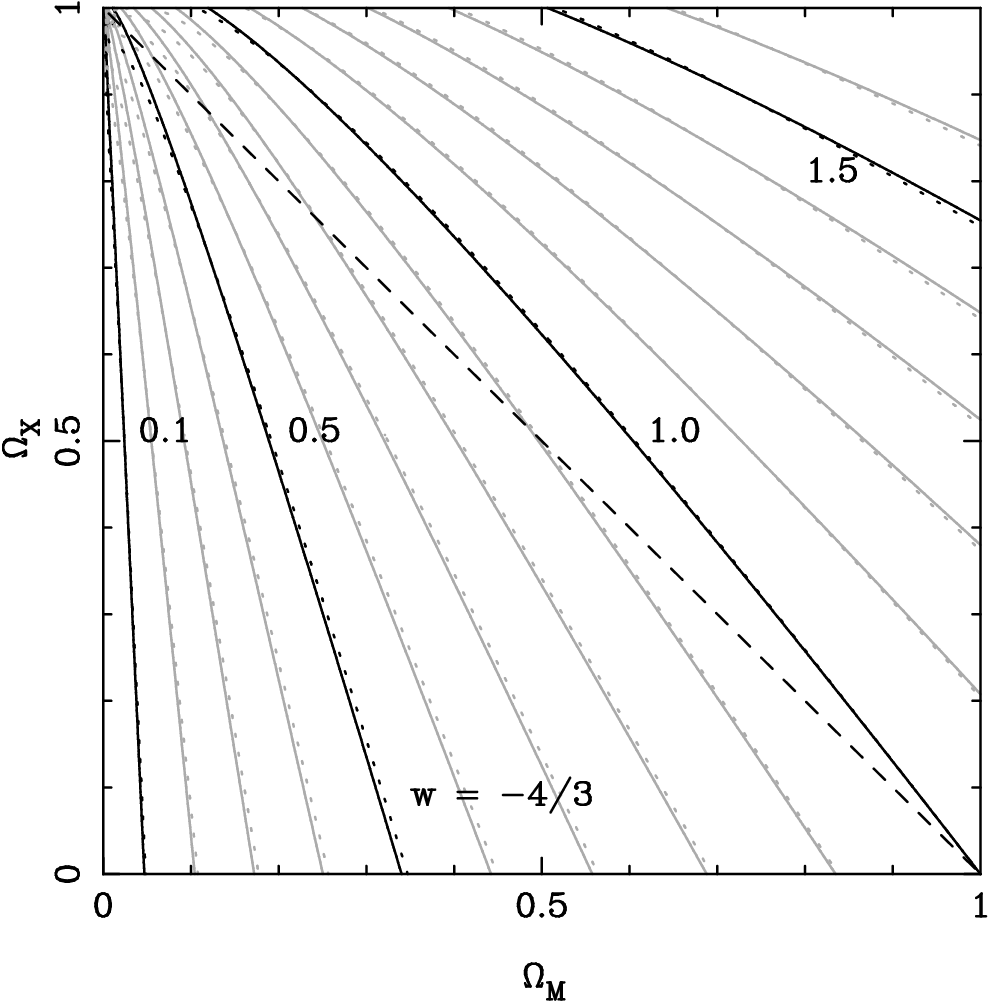}}
  \caption{Plots showing contours of constant linear growth factor to
  present day for a selection of cosmologies (grey and black
  lines). The growth factor is normalised to unity for $\Omega_M=1$,
  $\Omega_X=0$. Increasing $w$ from $-4/3$ to $-2/3$, decreases the
  growth factor leading to behaviour more like that of an open
  Universe. Dotted contours are for the fitting formula of
  Eq.~\ref{eq:Dapprox_quin}, together with
  Eqns.~\ref{eq:Dlin_alpha_fit} \&~\ref{eq:Dlin_A_fit}. The dashed
  line highlights flat cosmological models. \label{fig:lin_grow_w}}
\end{figure}

Eq.~\ref{eq:dlndda2} can easily be solved by numerical integration,
and we show contours of different linear growth factor to present day
in ($\Omega_M$,$\Omega_X$) space in Fig.~\ref{fig:lin_grow_w} for
constant $w=-2/3,-1, -4/3$. The linear growth factor is strongly
dependent on $w$, with $w>-1$ models behaving more like open Universes
than $w<-1$ models as the effect of the dark energy diminishes. For
the linear growth factor in a variety of cosmological models with time
dependent $w(a)$ see, for example, \citet{doran01,linder03}.

For $\Lambda$ cosmologies, the growing mode solution to this equation
is \citep{heath77}
\be
  D(a) = \frac{5\Omega_M}{2} E(a) \int^{a}_0\frac{da'}{[a'E(a')]^3},
    \label{eq:lin_growth_factor}
\ee
where $E(a)$ is given by Eq.~\ref{eq:friedmann_eq1}. Although this
integral can be easily solved numerically, it is common to use the
approximation of \citet{carroll}, which follows from work in
\citet{lightman90} and \citet{lahav91}
\begin{eqnarray}
  \nonumber
  \lefteqn{D(a) \simeq \frac{5\Omega_M(a)a}{2}
    \left[\Omega_M(a)^{4/7}-\Omega_\Lambda(a)\right.}\\
  & & \hspace{2.0cm} \left.+\left(1+\frac{\Omega_M(a)}{2}\right)
    \left(1+\frac{\Omega_\Lambda(a)}{70}\right) \right]^{-1}.
    \label{eq:Dapprox}
\end{eqnarray}

A general solution for the growing mode solution in dark energy
cosmologies, equivalent to Eq.~\ref{eq:lin_growth_factor}, has yet to
be found. However, for flat cosmological models, with constant $w$,
the solution can be written in terms of the hypergeometric function
$_2F_1$ \citep{silveira94}
\be
  D(a) = a\,_2F_1\left[-\frac{1}{3w},\frac{w-1}{2w},1-\frac{5}{6w},
	              -a^{-3w}\frac{1-\Omega_M}{\Omega_M}\right].
    \label{eq:Dhyp}
\ee

\begin{figure}
  \centering
  \resizebox{6.5cm}{!}{\includegraphics{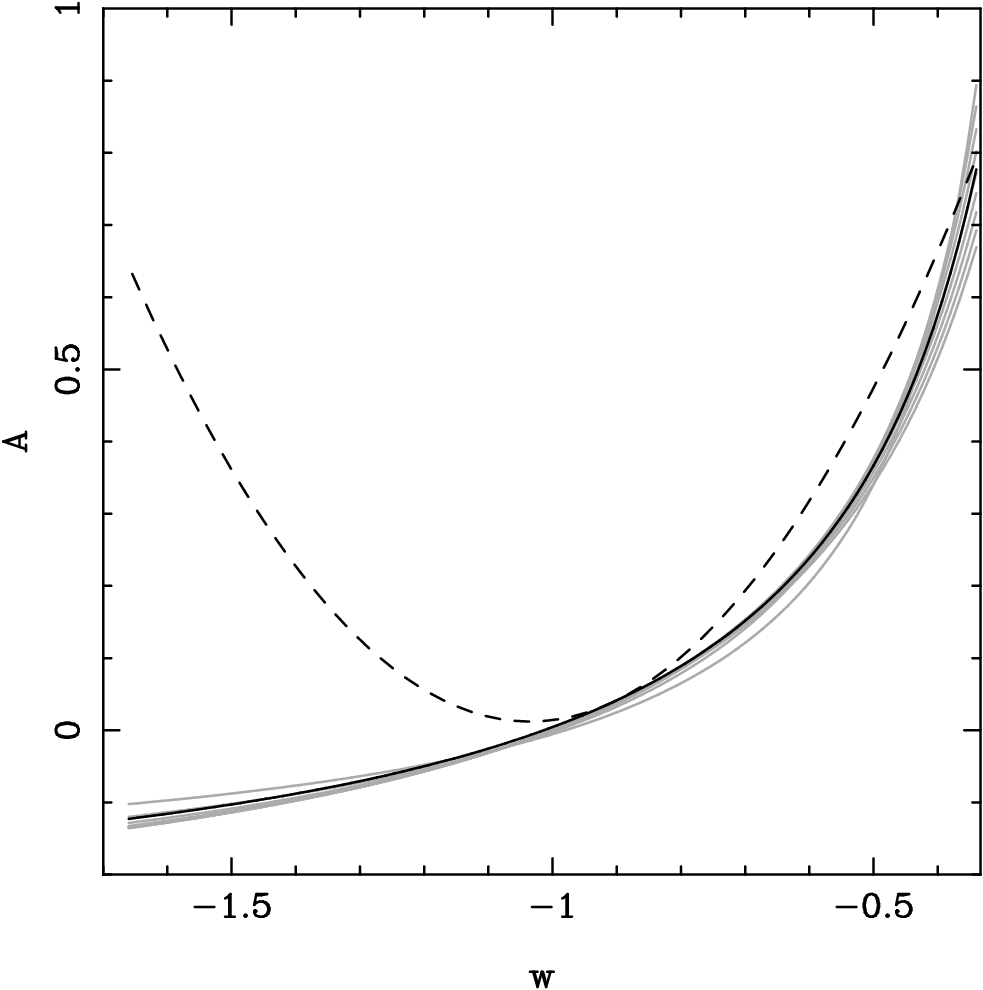}}
  \caption{Plot showing the true value of $\mathcal{A}$ in
  Eq.~\ref{eq:Dapprox_quin} as a function of $w$ for 9 flat
  cosmologies with $\Omega_M$ evenly spread between $0.1$ and $0.9$
  (grey lines). For comparison we plot the fitting formula of
  \citet{basilakos03} (dashed line) and for Eq.~\ref{eq:Dlin_A_fit} (black
  line). \label{fig:Dlin_A_fit}}
\end{figure}

Writing the growth index as
\be
  \frac{d\ln\delta}{d\ln a} = \Omega_M^\alpha(a),
\ee
\citet{wang98} use Eq.~\ref{eq:dlndda2} for the special case of flat
cosmologies to give 
\begin{displaymath}
  \alpha \simeq \frac{3}{5-w/(1-w)}
\end{displaymath}
\be
  \hspace{0.5cm}
    +\frac{3}{125}
    \frac{(1-w)(1-3w/2)}{(1-6w/5)^3}\left[1-\Omega_M(a)\right].
  \label{eq:Dlin_alpha_fit}
\ee
This led \citet{basilakos03} to extend the approximation of
\citet{carroll} given by Eq.~\ref{eq:Dapprox} to the case of $w\ne-1$
\begin{eqnarray}
  \nonumber
  \lefteqn{D(a) \simeq \frac{5\Omega_M(a)a}{2}
    \left[\Omega_M(a)^\alpha-\Omega_X(a)\right.}\\
  & & \hspace{2.0cm} \left.+\left(1+\frac{\Omega_M(a)}{2}\right)
    \left(1+\mathcal{A}\Omega_X(a)\right) \right]^{-1},
    \label{eq:Dapprox_quin}
\end{eqnarray}

with $\alpha$ given by Eq.~\ref{eq:Dlin_alpha_fit}, and
$\mathcal{A}\simeq1.742+3.343w+1.615w^2$. In Fig.~\ref{fig:Dlin_A_fit}
we plot the $\mathcal{A}$ values required to match
Eq.~\ref{eq:Dapprox_quin} to the true linear growth factor (given by
Eq.~\ref{eq:Dhyp}), for flat cosmological models with
$0.1<\Omega_M<0.9$ as a function of $w$ (grey lines). The fit of
\citet{basilakos03} is shown by the dashed line. This is a poor fit for
$w<-1$, so instead, we propose
\be
  \mathcal{A} = \frac{-0.28}{w+0.08}-0.3,
    \label{eq:Dlin_A_fit}
\ee
shown by the black line in Fig.~\ref{fig:Dlin_A_fit}.

Eq.~\ref{eq:Dlin_A_fit} has been determined by fitting to flat
cosmological models with $0.1<\Omega_M<0.9$. For non-flat models, the
approximation of Eq.~\ref{eq:Dapprox_quin} remains a good fit. In
Fig.~\ref{fig:lin_grow_w}, the dotted contours are for the fitting
formula of Eq.~\ref{eq:Dapprox_quin}, together with
Eqns.~\ref{eq:Dlin_alpha_fit} \&~\ref{eq:Dlin_A_fit}, compared with
the true value of $D_0$ given by the solid contours. The fitting
formula fails for $\Omega_M<<0.1$, but for $\Omega_M>0.1$, the maximum
error (with $0<\Omega_X<1$) is 3.8\% for $w=-4/3$, 2.6\% for $w=-1$
and 5.1\% for $w=-2/3$. For comparison, the fitting formula of
\citet{carroll} given by Eq.~\ref{eq:Dapprox} is accurate to 2.1\% for
$w=-1$ over this range of $\Omega_M$.

\section{the critical density for collapse of spherical perturbations}
  \label{sec:dc}

We now calculate the critical overdensity for collapse of homogeneous
spherical perturbations at present day in a homogeneous dark energy
background. The method adopted is a development of that in
\citet{percival00}, where the critical overdensity in $\Lambda$
cosmologies was calculated. Solution schemes for an Einstein-de Sitter
cosmology \citep{gunn72}, for open cosmologies \citet{lacey93} and for
flat $\Lambda$ cosmologies \citep{eke96} were summarised in
\citet{kitayama96}. See also the solution scheme for $\Lambda$
cosmologies used by \citet{barrow84,barrow93}.

As in Section ~\ref{sec:linear}, we consider two spheres containing
equal amounts of material: one of background material with radius $a$,
and one of radius $a_p$ with a homogeneous change in overdensity. If
the dark energy component is negligible, as at early times for
$lim_{a\to0}w(a)<-1/3$,
\be
  \left(\frac{da}{d(H_0t)}\right)^2=\frac{\Omega_M}{a}+\epsilon,
  \label{eq:friedmann1}
\ee
where $\epsilon$ is allowed to take any real value. For the
background, $\epsilon=\Omega_K\equiv(1-\Omega_M-\Omega_X)$ is the
standard curvature constant. The matter density $\Omega_M$ is the same
for both the perturbation and the background as the two spherical
regions contain the same mass. Following \citet{percival00}, a series
solution for $a(H_0t)$ in the limit $H_0t\to0$ can be obtained given
by $a=\alpha (H_0t)^{2/3}+\beta (H_0t)^{4/3}+O[(H_0t)^{6/3}]$, where
\be
  \alpha = \left(\frac{9\Omega_M}{4}\right)^{1/3}, \,\,\,
  \beta  = \frac{3\epsilon}{20}\left(\frac{12}{\Omega_M}\right)^{1/3}.
    \label{eq:ab}
\ee
Using the fact that the spheres contain equal mass,
\be
  \lim_{H_0t\to0}\delta(H_0t)=
    \frac{3}{\alpha}(\beta-\beta_p)(H_0t)^{2/3}+O[(H_0t)^{4/3}]. 
  \label{eq:limits}
\ee
Substituting Eq.~\ref{eq:ab} into Eq.~\ref{eq:limits} gives that
\begin{displaymath}
  \lim_{H_0t\to0}\delta(H_0t)=
\end{displaymath}
\be
  \hspace{1cm}
    \frac{3}{5}\left(\frac{3}{2\Omega_M}\right)^{2/3}
    \left[(1-\Omega_M-\Omega_X)-\epsilon_p\right]
    (H_0t)^{2/3}.
  \label{eq:ktolimd}
\ee

We now extrapolate this limiting behaviour to present day, using the
linear growth factor to extrapolate the normalisation of the density
field from early times. This gives
\be
  \delta_c = D_0\lim_{H_0t\to0}\left[\frac{\delta(t)}{D(t)}\right],
  \label{eq:delta_c_extrap}
\ee
where $D_0$ is the linear growth factor to present day. Given the
normalisation of $D(t)$ adopted in Section~\ref{sec:linear},
$\lim_{H_0t\to0}D(H_0t)=\alpha(H_0t)^{2/3}$, where $\alpha$ is given
by Eq.~\ref{eq:ab}. We can therefore write $\delta_c$ as
\be
  \frac{\delta_c}{D_0}= \frac{3}{5\Omega_M}
    \left[(1-\Omega_M-\Omega_X)-\epsilon_p\right].
  \label{eq:delta_c1}
\ee

This formula relates the linearly extrapolated overdensity $\delta_c$
to the initial curvature of the perturbation for cosmological models
in which the dark energy becomes negligible as $H_0t\to0$. 

\begin{figure}
  \centering
  \resizebox{5.6cm}{!}{\includegraphics{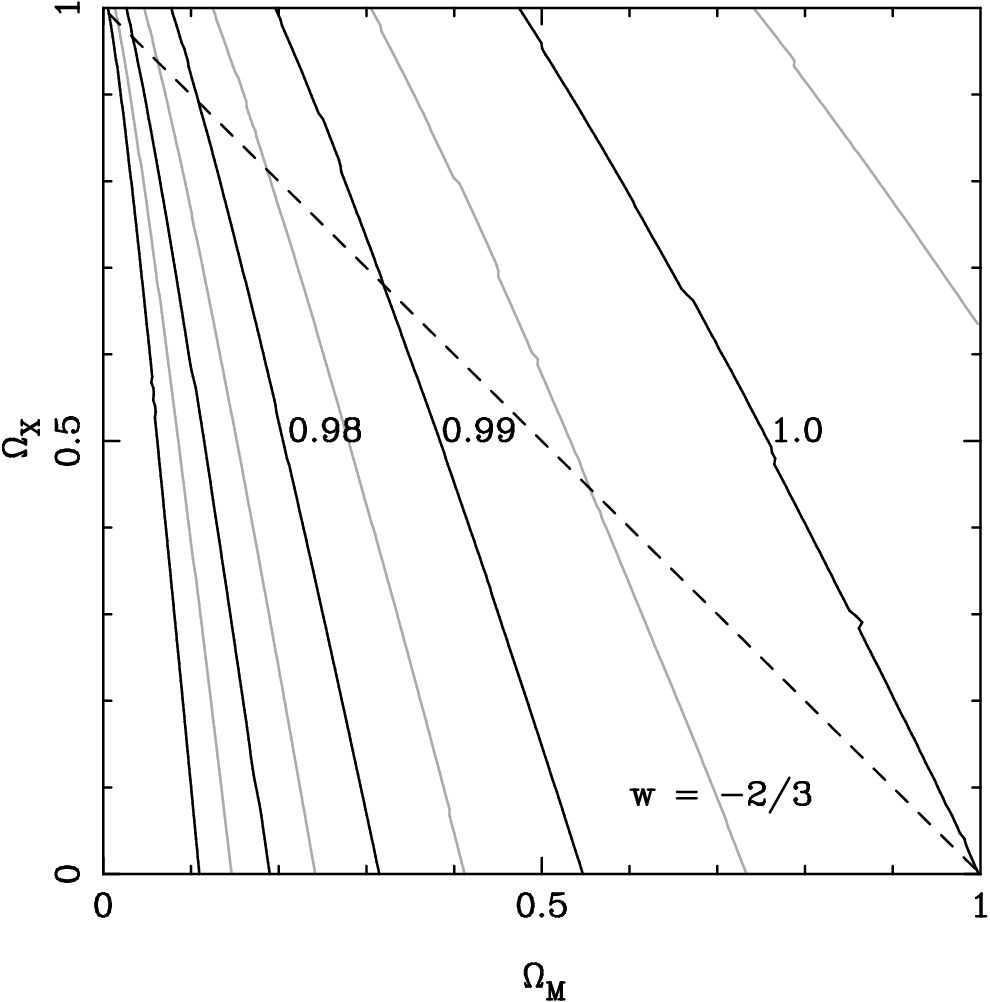}}
  \resizebox{5.6cm}{!}{\includegraphics{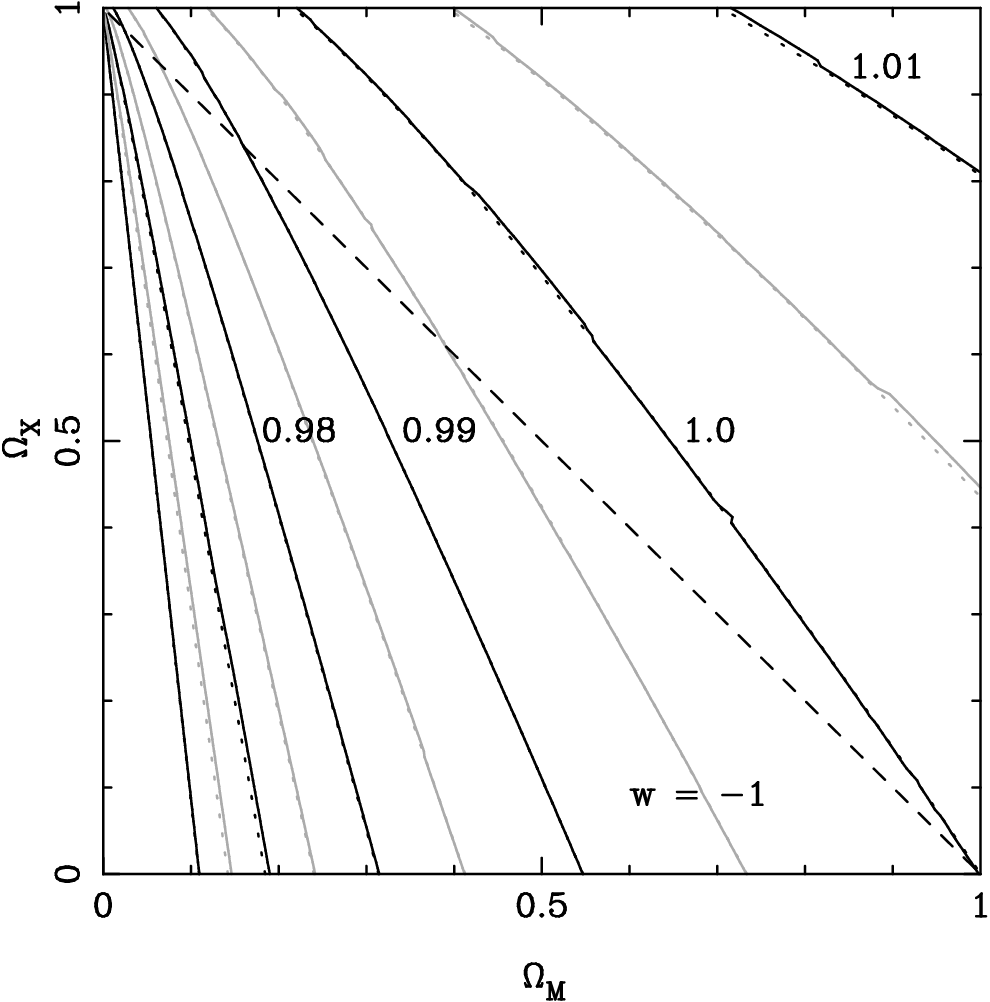}}
  \resizebox{5.6cm}{!}{\includegraphics{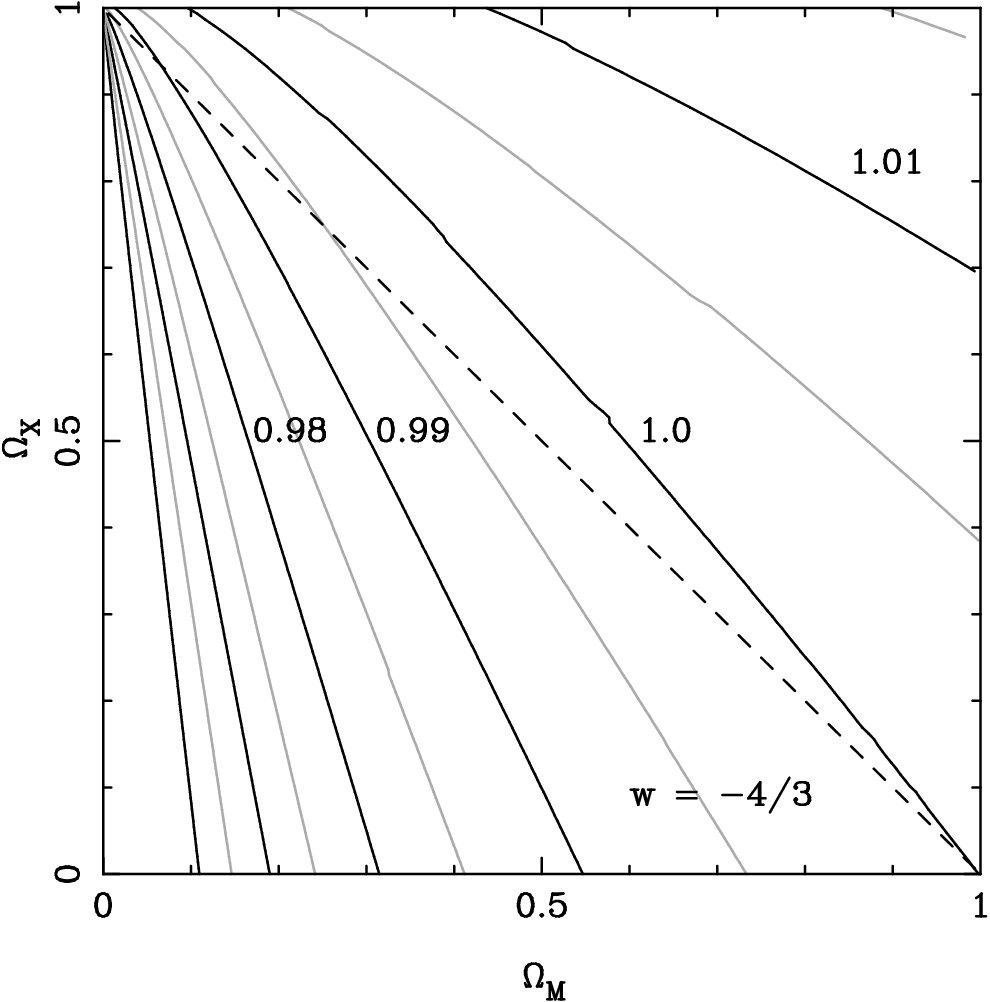}}
  \caption{Plots showing contours of constant critical overdensity for
  a selection of cosmologies (grey and black lines) with
  constant $w$. Contours are plotted as a ratio of $\delta_{\rm EdS}$
  as given by Eq.~\ref{eq:delta_eds} from $0.96\delta_{\rm EdS}$ to
  $1.02\delta_{\rm EdS}$ at intervals of $0.005\delta_{\rm EdS}$. For
  the $w=-1$ cosmology we also plot the values determined by
  accurately solving Eq.~\ref{eq:friedmann2} using standard Elliptical
  integrals (dotted lines). The difference between solid and dotted
  lines are caused by the numerical integration of
  Eqns~\ref{eq:num_int1}, \ref{eq:num_int2} \& \ref{eq:num_int3}. The
  dashed line highlights flat cosmological models (as in
  Fig.~\ref{fig:lin_grow_w}). \label{fig:dc_w}}
\end{figure}

\subsection{case 1: cosmological constant} \label{sec:delc_lambda}

For $\Lambda$ cosmologies, the curvature of the perturbation (and
energy) is conserved through the Friedmann equation
\be
  \left(\frac{da_p}{d(H_0t)}\right)^2=\frac{\Omega_M}{a_p}+\epsilon_p
    + \Omega_\Lambda a_p^2.
  \label{eq:friedmann2}
\ee

This equation defines a mini-cosmology, so using the methodology of
Section~\ref{sec:dynamics}, the requirement for collapse can be seen
to be $-\epsilon_p^3>27\Omega_\Lambda\Omega_M^2/4$ (compare with
Eq.~\ref{eq:omega_critical}). Given a perturbation that collapses, the
time taken can be found by integrating Eq.~\ref{eq:friedmann2}, which
can be reduced to a function of standard Elliptical integrals.

If $\Omega_\Lambda=0$, Eq.~\ref{eq:friedmann2} can be solved
analytically and, for an Einstein-de Sitter cosmology, the critical
overdensity for collapse at present day reduces to
\be
  \delta_{\rm EdS} \equiv \frac{3}{20}(12\pi)^{2/3} \simeq 1.686,
    \label{eq:delta_eds}
\ee
which was first derived by \citet{gunn72}.

\subsection{case 2: general $w(a)$ cosmologies} \label{sec:delc_quinn}

When the dark energy does not cluster, the energy within a spherical
perturbation is not conserved (a comprehensive discussion of this is
given in \citealt{weinberg03}). Although this means that we cannot use
Eq.~\ref{eq:friedmann2} to determine the behaviour of the
perturbation, we can still use the cosmology equation
(Eq.~\ref{eq:cosmo}). \citet{wang98} provide a boundary value problem
for solving this second order differential equation, setting the
boundary at the turn-around time. However, given the relatively
straightforward initial evolution considered above, it is far simpler
to set up an initial value problem (in either $a_p$ or $\delta$) to
determine the subsequent behaviour. In a recent paper, \citet{chiba05}
have also determined the limiting initial conditions for the
differential equations, working directly from the cosmology
equation. However, it is perhaps more intuitive to link the critical
overdensity to the initial curvature of the perturbation
(Eq.~\ref{eq:delta_c1}).

To see how this proceeds from the derivation above, suppose that we
know $\delta_c$. Then Eq.~\ref{eq:delta_c1} can be used to find the
initial curvature $\epsilon_p$. The initial conditions, $a_p$,
$\delta$, and $da_p/dt$ at some small, but finite $a$ can be obtained
by substituting $\epsilon_p$ into Eqns~\ref{eq:friedmann1}
\&~\ref{eq:ktolimd}.

Given these initial conditions, the evolution of the perturbation is
uniquely specified by three equations: the cosmology equation for the
perturbation and the background, and the Friedmann equation for the
background. For completeness we repeat these equations here.
\be
  \frac{1}{a_p}\frac{d^2a_p}{d(H_0t)^2}=-\frac{1}{2}
  \left[\Omega_Ma_p^{-3}+(3w(a)+1)\Omega_Xa^{f(a)}\right],
  \label{eq:num_int1}
\ee
\be
  \frac{1}{a}\frac{d^2a}{d(H_0t)^2}=-\frac{1}{2}
  \left[\Omega_Ma^{-3}+(3w(a)+1)\Omega_Xa^{f(a)}\right],
  \label{eq:num_int2}
\ee
\be
  \frac{1}{a^2}\left[\frac{da}{d(H_0t)}\right]^2=
    \Omega_Ma^{-3}+\Omega_Ka^{-2}+\Omega_Xa^{f(a)}.
  \label{eq:num_int3}
\ee

By setting the initial value for the solution of these equations, we
avoid any issues to do with symmetry: for general $w(a)$ cosmologies,
not only is Eq.~\ref{eq:friedmann2} invalid, but additionally the
evolution of $a_p$ in time is no longer symmetric about the point of
maximum expansion, because the effect of the dark energy is not
symmetric about this point. This asymmetry is automatically accounted
for by solving these equations from an initial value. In fact, this
asymmetry leads to potentially interesting behaviour for the spherical
perturbations: given the right initial conditions, perturbations can
start to recollapse (turn-around), but then the repulsive force of the
dark energy can cause re-expansion \citep{chiba05}.

In Fig.~\ref{fig:dc_w}, we show contours of constant $\delta_c$ as a
function of $\Omega_M$ and $\Omega_X$ for constant $w=-4/3,-1,
-2/3$. Contours are plotted as a function of $\delta_{\rm EdS}$ as
given by Eq.~\ref{eq:delta_eds}. As $\Omega_X\to0$, the solutions
asymptote towards the open Universe values calculated by
\citet{lacey93}. For $\Omega_X>0$, we see that decreasing $w$ from
$-2/3$ to $-4/3$ increases the effect of the dark energy. As for the
linear growth factor, $w>-1$ cosmologies behave more like open
cosmologies than those with $w<-1$: decreasing $w$ increases the
importance of the dark energy for calculating $\delta_c$. The critical
overdensities for collapse in a variety of cosmologies with varying
$w(a)$ are compared in \citet{mota04}.

\section{the evolution of the critical overdensity}  \label{sec:dc_evol}

The evolution of the critical overdensity for collapse $\delta_c(a)$,
is usually defined as follows: for a cosmological model with
parameters $\Omega_M$ \& $\Omega_X$, $\delta_c(a)$ gives the
overdensity for a perturbation that collapses at scale factor $a$,
normalised at present day. For example if we had a density field (and
associated power spectrum) normalised at present day, then
$\delta_c(a)$ (where $a$ does not necessarily equal 1) relates to
spherical perturbations in this density field that collapse at scale
factor $a$. $\delta_c$ is the particular case for perturbations that
collapse at present day: perturbations that collapse earlier obviously
have to be significantly more overdense. To calculate $\delta_c(a)$,
when integrating the Friedmann equation (Eq.~\ref{eq:friedmann2}) for
$\Lambda$ cosmologies, or numerically integrating
Eqns.~\ref{eq:num_int1}, \ref{eq:num_int2} \& \ref{eq:num_int3} we
would look for collapse at scale factor $a$ rather than $a=1$. 

The rather weak evolution of the critical overdensity as a function of
cosmological model (Fig.~\ref{fig:dc_w}), means that the linear growth
factor can be used to approximate $\delta_c(a)$: If $\delta_c$ is
constant along a particular cosmological track then the evolution of
$\delta_c(a)$ is purely driven by $D(a)^{-1}$. The only change in
$\delta_c(a)$ between two collapse times is caused by the change in 
overall normalisation of the field, which can be seen by considering
Eq.~\ref{eq:delta_c_extrap} for two cosmologies along a single
cosmological track. The most obvious choice for the normalisation is
$\delta_{\rm EdS}$, so the approximation will be correct in the limit
as $a\to0$,
\be
  \delta_c(a) \simeq \frac{D_0}{D(a)} \delta_{\rm EdS}.
  \label{eq:dc_approx}
\ee 

\begin{figure}
  \centering
  \resizebox{6.5cm}{!}{\includegraphics{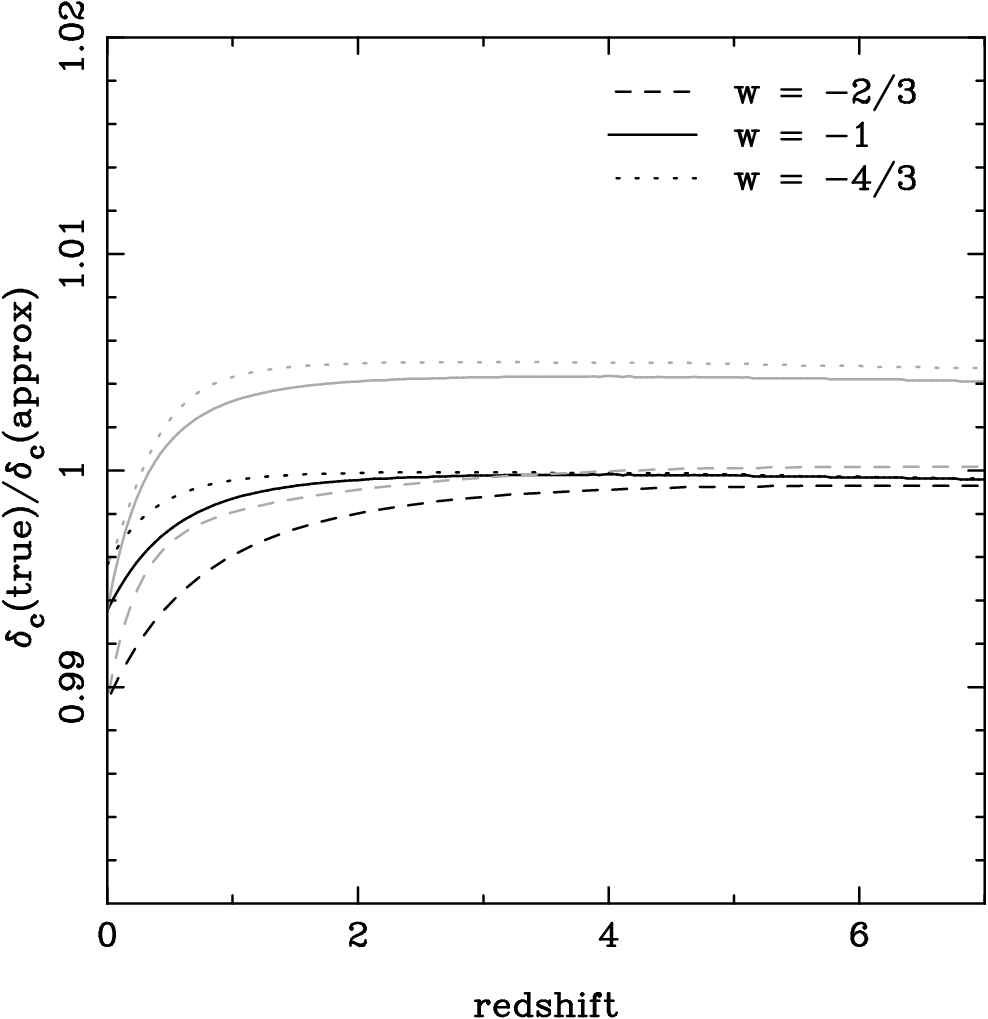}}

  \caption{Plot showing the ratio between the critical overdensity for
  collapse $\delta_c(a)$ and the approximation given by
  Eq.~\ref{eq:dc_approx}, as a function of redshift for three
  different cosmological models (black lines). $\Omega_M=0.3$ and
  $\Omega_X=0.7$ are assumed, while $w$ is assumed to be constant with
  $w=-2/3,-1$, or $-4/3$. The grey lines shows the ratio between the
  true critical overdensity for collapse and the approximation, where
  the linear growth factor has itself been approximated by
  Eq.~\ref{eq:Dapprox_quin}. The error in using the Einstein-de Sitter
  critical overdensity is of the same order as the error in the
  approximation of using the fitting formula of
  Eq.~\ref{eq:Dapprox_quin}. \label{fig:dc_approx}}
\end{figure}

In order to demonstrate this approximation, in
Fig.~\ref{fig:dc_approx}, we plot the ratio between the true critical
overdensity $\delta_c(a)$ to the approximation given by
Eq.~\ref{eq:dc_approx} for three cosmological models with
$\Omega_M=0.3$, $\Omega_X=0.7$, but with constant $w=-2/3,-1,-4/3$. As
$a\to0$, the approximation becomes more accurate. At $a=1$ (redshift
$0$), the ratio is the same as the value of the surface contoured in
Fig.~\ref{fig:dc_w}. We also plot the approximate value of
$\delta_c(a)$ calculated from Eq.~\ref{eq:dc_approx}, but with $D(a)$
and $D_0$ calculated using the approximation given by
Eq.~\ref{eq:Dapprox_quin}, Eq.~\ref{eq:Dlin_A_fit}
\&~Eq.~\ref{eq:Dlin_alpha_fit} (grey lines). The error from using the
approximation of Eq.~\ref{eq:dc_approx} is of the same order as that
from assuming the approximation to the linear growth factor for these
cosmologies.

\section{the mass function}  \label{sec:mass_func}

The usefulness of the spherical model was emphasised when Press \&
Schechter considered smoothing the initial density field to determine
the relative abundances of perturbations on different scales
(\citealt{ps}: PS). When combined with the critical overdensity for
collapse this provided a statistical model for the formation of
structure in the Universe: smoothing the fluctuations leads to the
masses of collapsed objects, while the spherical perturbation model
gives the epoch of collapse for those perturbations that are
sufficiently dense. Obviously such a simple model will fail in detail,
particularly given the known complexities of asymmetrical
gravitational collapse, and numerical simulations have now quantified
these problems \citep{st99, jenkins2001}. However PS theory has been
incredibly successful and arguably still provides key insight into the
processes at work in structure formation.

Two alternative formalisms are often considered for the collapse of
perturbations in PS theory
\begin{enumerate}
  \item The overdensity field is assumed to grow with the linear
  growth factor, and when perturbations reach the critical overdensity
  they are said to have collapsed.
  \item Each overdense region is considered to be spherical and its
  collapse time is calculated as in Section~\ref{sec:delc_lambda}.
\end{enumerate}
Given the discussion in Section~\ref{sec:dc_evol}, it is easy to see
that the first formalism, corresponding to a growing field, matches
the approximation to $\delta_c(a)$ using the linear growth factor
given by Eq.~\ref{eq:dc_approx}. The second formalism, which is
adopted in the following analysis, corresponds to using the correct
$\delta_c(a)$ for the spherical model.

\citet{ma99} considered the effect of quintessence on the mass
transfer function. They provided fitting formulae for the ratio
between the quintessence and $\Lambda$ cosmologies. However, if the
dark energy only clusters on very large scales, the transfer function
is only altered on these scales. If the power spectrum is normalised
to $\sigma_8$ (the rms density fluctuation on scales of $8\hompc$),
then the scales usually of interest are not affected \citep{lokas04}.

\begin{figure}
  \centering
  \resizebox{6.5cm}{!}{\includegraphics{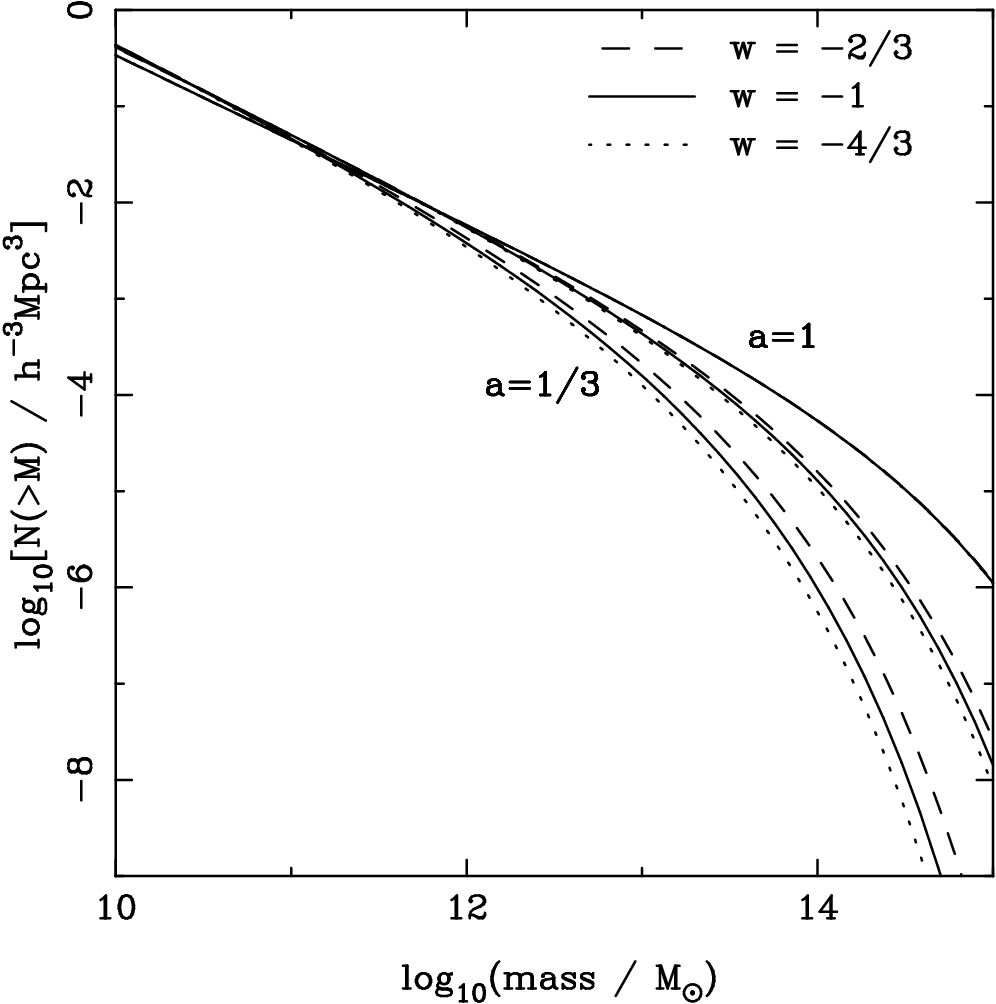}}

  \caption{Plot showing the predicted mass function calculated using
the fitting formula of \citet{st99} calculated for $\Omega_M=0.3$,
$\Omega_X=0.7$ for three different values of $w$, and at three epochs
corresponding to $a=1/3, 1/2, 1$. Because the power spectrum is
normalised at present day, and $\delta_c$ is only weakly dependent on
cosmology, then there is little difference between the predicted mass
functions for $a=1$. As we go further back in time the difference
becomes more severe because of the differing linear growth
factors. \label{fig:mass_function_w}}
\end{figure}

To demonstrate the effect of the dark energy equation of state on the
mass function, we plot the cumulative mass function $N(>M)$,
calculated using the numerical fit of \citet{st99} for 3 different
cosmologies and 3 different epochs in
Fig.~\ref{fig:mass_function_w}. To determine the power spectrum we
have used the fitting formulae of \citet{eh98} with $\Omega_M=0.3$,
$\Omega_b/\Omega_M=0.15$, $\sigma_8=0.9$, $n_s=1$. The critical
overdensity for collapse for $w=-2/3,-1$ \& $-4/3$ and $\Omega_X=0.7$
was then calculated for $a=1/3,1/2,1$, corresponding to redshifts
$z=2,1,0$. As expected, because the critical overdensity for collapse
is only weakly dependent on cosmological parameters
(Fig.~\ref{fig:dc_w}), at $a=1$ (the epoch at which the power spectrum
is normalised) we see very little difference in the predicted mass
functions for different cosmologies. If the normalisation of the power
spectrum had been constrained at a different epoch (for example by CMB
fluctuations), then this would not be correct. The evolution of the
mass function is strongly dependent on $w$ because of the effect on
the evolution of $\delta_c(a)$ (Section~\ref{sec:dc_evol}) through the
linear growth factor (Fig.~\ref{fig:lin_grow_w}). Consequently,
determining the mass function at redshifts other than that used to
normalise the power spectrum offers a stronger possibility of
measuring $w(a)$. We will discuss the evolution of structure growth
further in the next Section.

In order to compare with the observed cluster counts, we must convert
from comoving position to observed angular position and redshift. The
number of sources with mass $>M$ per unit solid angle in a redshift
slice $dz$ is given by 
\be
  N(>M, \Delta z) = \int_z^{z+\Delta z}\,dz\, N(>M) d_{\rm prop}^2 
    \frac{dd_{\rm prop}}{dz},
\ee
where the proper distance $d_{\rm prop}$ is given by
\be
  d_{\rm prop}= \frac{c}{H_0}\int_a^1\,\frac{da}{a^2E(a)}.
\ee
This correction to the mass function reduces the significance of $w$
at low redshifts for masses of order $10^{14}\msun$ \citep{solevi05}.

\section{rate of structure growth}  \label{sec:rate_grow}

As discussed in the previous Section, the present day mass function
does not provide a good test of $w$, although its evolution in time
does. The Press-Schechter mass function (and the fitting formula of
\citealt{st99}) is independent of epoch when written as a function of
$\nu=\delta_c/\sigma_M$, where $\sigma_M$ is the rms fluctuation in
the initial overdensity field on a scale corresponding to mass
$M$. Consequently, it is easy to see that the evolution of the mass
function is dependent on the rate at which $\delta_c$ changes,
$d\delta_c(a)/da$. It is this quantity that we are testing by
comparing mass functions at different epochs.

Extended Press-Schechter theory uses the smoothed initial overdensity
field to create an analytic model for the build-up of structure
\citep{bond91,bower91,lacey93}, in addition to providing a statistical
model for the mass function. Within this extension to the theory, the
rate at which structures grow is again driven by $d\delta_c(a)/da$
\citep{percival99,percival00,cole00,miller05}. Appendix~A of
\citet{miller05} showed that the approximation of
Eq.~\ref{eq:dc_approx} can be easily extended to $d\delta_c(a)/da$ for
$\Lambda$ cosmologies. 

\begin{figure}
  \centering
  \resizebox{6.5cm}{!}{\includegraphics{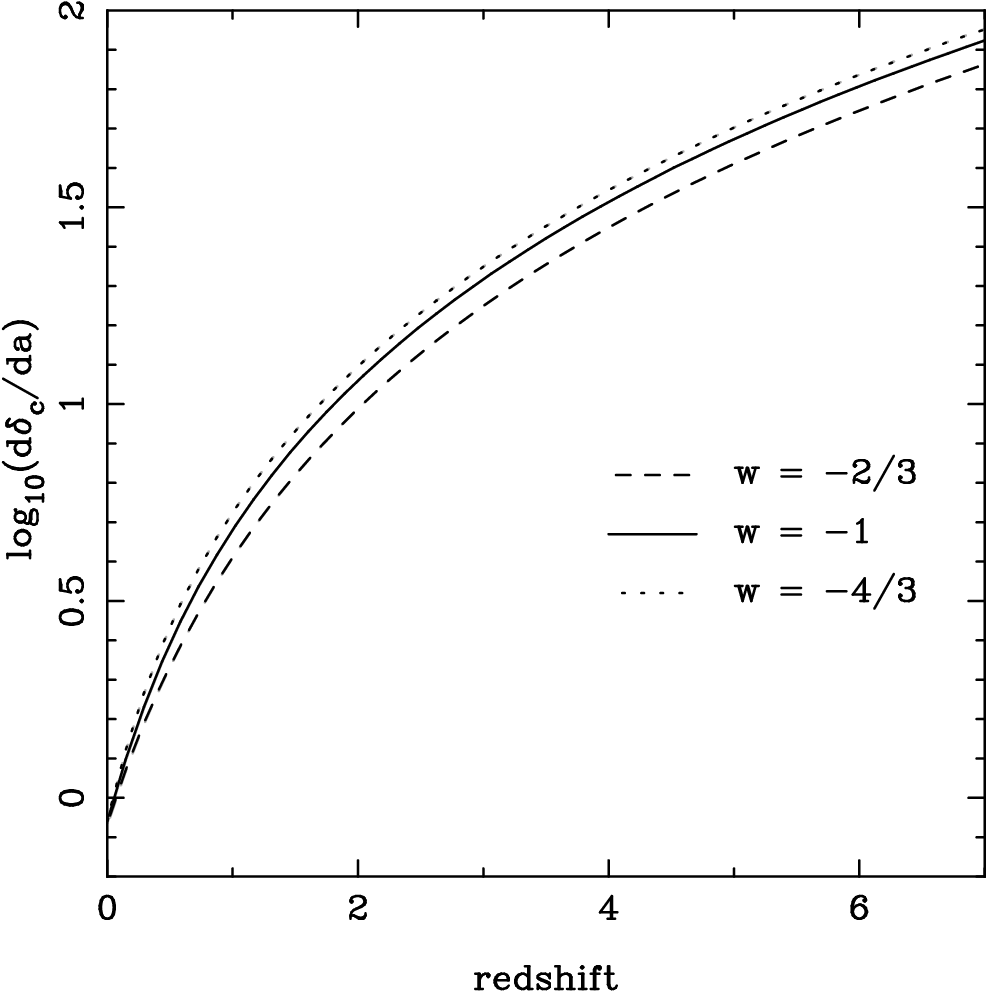}}

  \caption{Plot showing the derivative of the critical overdensity for
  collapse with respect to the scale factor $d\delta_c(a)/da$ (black
  lines) and the approximation calculated from Eq.~\ref{eq:dc_approx}
  (grey lines), as a function of redshift for three different
  cosmological models. $\Omega_M=0.3$ and $\Omega_X=0.7$ are assumed,
  while $w=-4/3,-1$, or $-2/3$. The approximation is so good that
  the grey lines are almost completely hidden behind the black lines.
  \label{fig:ddcda}}
\end{figure}

Here, we extend this analysis to consider $d\delta_c(a)/da$ for
$w\ne-1$ cosmologies. In Fig.~\ref{fig:ddcda} we show
$d\delta_c(a)/da$ for three flat cosmological models with
$\Omega_M=0.3$, and $w=-4/3,-1$, or $-2/3$ calculated using the
derivation of Section~\ref{sec:delc_quinn} (black lines) and the fit
of Eq.~\ref{eq:dc_approx} (grey lines). In fact, the fit is so good
that the grey lines are hardly visible in this plot. As expected
$d\delta_c(a)/da$ is strongly dependent on the dark energy equation of
state.

\section{virialisation}  \label{sec:virial}

The inhomogeneous nature of true perturbations means that they do not
collapse to singularities, but instead stabilise at finite size. For
$\Lambda$ cosmologies it is possible to use energy considerations to
determine the final radius and density of the virialised perturbation
\citep{lahav91}. The extension of this analysis to more general dark
energy cosmologies has been considered by a number of authors
\citep{wang98,horellou05,maor05}.

Within a perturbation, the potential energy due to the matter $U_G$ and
dark energy $U_X$ are given by \citep{horellou05,maor05}
\be
  U_G = -\frac{3GM^2}{5R},\,\,\, 
  U_X = [1+3w(a)]\frac{4\pi GM}{10}\rho_X R^2.
  \label{eq:energy}
\ee
These can be calculated from the Poisson Equation with pressure
term. Note that there is some confusion in the literature about the
exact form of $U_X$ and the $[1+3w(a)]$ term has sometimes been
neglected in the past, although it is included in more recent work
\citep{battye03,horellou05}. Even without this term, the discussion
below about the lack of energy conservation within the perturbation
remains valid, although the numerical results will obviously change.

For dark energy cosmologies the virial theorem, that a system with
potential energy $U\propto R^p$ virialises with temperature $T=pU/2$,
holds and $T=-\frac{1}{2}U_G+U_X$ at the epoch of
virialisation. Assuming conservation of total energy between
turn-around at $a_{\rm ta}$ and virialisation $a_{\rm vir}$ gives
\be
  U_G(a_{\rm ta}) + U_X(a_{\rm ta}) = \frac{1}{2}U_G(a_{\rm vir}) +
    2U_X(a_{\rm vir}).
  \label{eq:totenergy}
\ee
Substituting the definitions given by Eq.~\ref{eq:energy} leads to
\citep{wang98,horellou05,maor05} 
\be
 \left(1-[1+3w(a_{\rm ta})]\frac{q}{2}\right)x 
    + [1+3w(a_{\rm vir})]\frac{qx^3}{y}
    =\frac{1}{2},
  \label{eq:r_ratio_const_energy_ta}	 
\ee
where 
\begin{eqnarray}
\nonumber
 x & = & \frac{R_{\rm vir}}{R_{\rm ta}},\\
\nonumber
 q & = & \frac{\Omega_X(a_{\rm ta})}
              {[1+\delta(a_{\rm ta})]\Omega_M(a_{\rm ta})},\\
 y & = & \frac{\rho_X(a_{\rm ta})}{\rho_X(a_{\rm vir})} 
     = \frac{a_{\rm ta}^{\,\,f(a_{\rm ta})}}{a_{\rm vir}^{\,\,f(a_{\rm vir})}}.
\end{eqnarray}
Here $q$ gives the ratio of the dark energy density to the matter
density in the perturbation at turn-around. For $w=-1$, $y=1$ and
Eq.~\ref{eq:r_ratio_const_energy_ta} reduces to the formula of
\citet{lahav91}. For $w=-1/3$ or $\Omega_X=0$, we find $x=1/2$.

The above derivation was based on the assumption that total energy is
conserved for the perturbation between turn-around and
virialisation. This is a common assumption in the literature for
cosmologies in which the dark energy remains homogeneous on the scales
of the perturbations
(e.g. \citealt{wang98,weinberg03,battye03,horellou05}). Even in work
that considers possible dark energy clustering, energy conservation
has previously been assumed in the homogeneous case (see the
discussion leading to equation 26 of \citealt{maor05}). However, the
potential energy of the matter due to the presence of the dark energy
$U_X$ ($U_{12}$ in the notation of \citealt{maor05}) is dependent on
the dark energy density in the perturbation $\rho_X(a)$. The evolution
of this density lies with the background, rather than the
perturbation, so the total energy is not expected to be conserved:
this is why we could not write down a Friedmann equation for the
perturbation in Section~\ref{sec:dc}. Obviously, if the total energy
of the perturbation between turn-around and virialisation is not
conserved, then Equation~\ref{eq:totenergy} does not hold.

\begin{figure}
  \centering
  \resizebox{6.5cm}{!}{\includegraphics{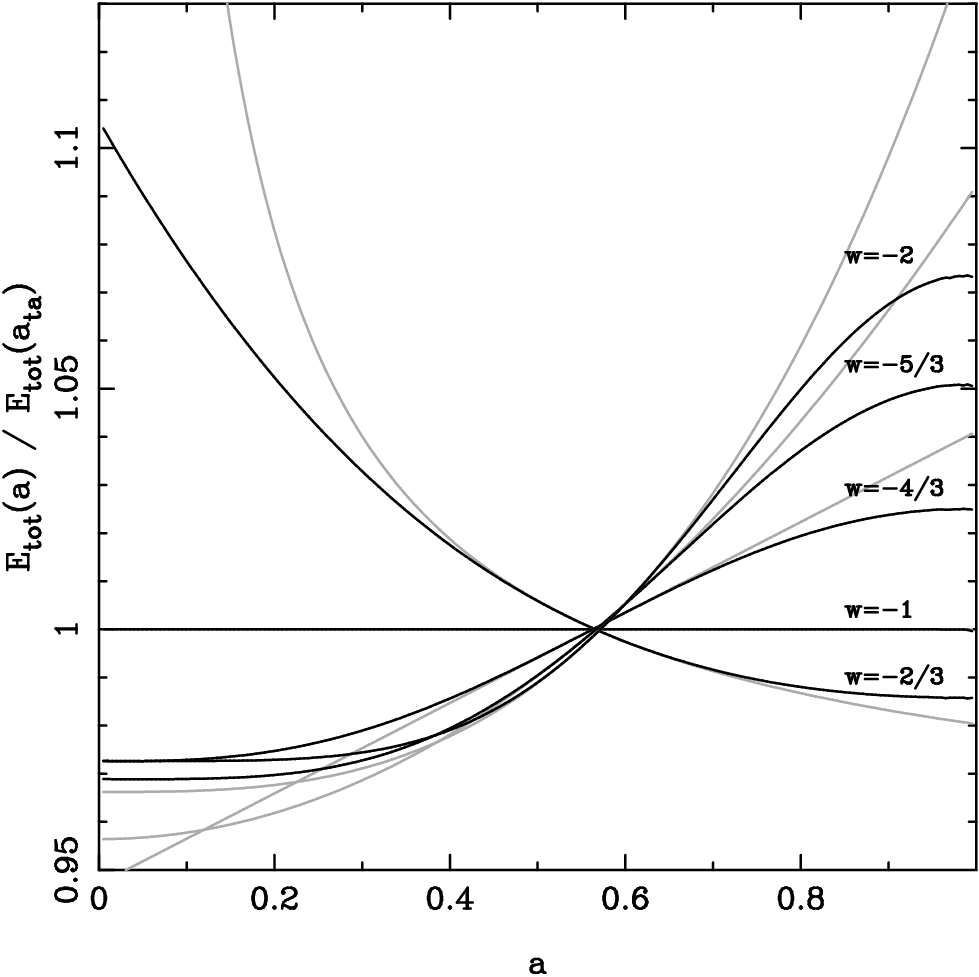}}

  \caption{The total energy in an ideal homogeneous spherical
  perturbation (as modelled in Section~\ref{sec:dc}), undergoing
  collapse to a singularity at $a=1$, relative to the energy at
  turn-around (black lines). $\Omega_M=0.3$ and $\Omega_X=0.7$ are
  assumed, with constant $w$ at the values shown. For general $w$, the
  total energy is not conserved. If the perturbations had remained at
  the turn-around size throughout their evolution, then the total
  energies would have evolved along the grey lines. The actual
  evolution in total energy lies between the line of constant total
  energy and the line of constant size for each
  cosmology. \label{fig:tot_energy_evol_w}}
\end{figure}

In fact, we can consider two extreme situations. In the first, as
above, the total energy at $a_{\rm vir}$ is the same as at $a_{\rm
ta}$. In the second, we assume that the perturbation did not change
size between the two epochs. Following the second assumption, the
total energy in the system at $a_{\rm vir}$ would have been altered
from that at $a_{\rm ta}$ by the change in the dark energy potential
$U_{X,1}-U_{X,2}$, where
\begin{eqnarray}
  U_{X,1} & = & 
    [1+3w(a_{\rm ta})]\frac{4\pi GM}{10}\rho_X(a_{\rm ta}) R_{\rm ta}^2,\\
  U_{X,2} & = & 
    [1+3w(a_{\rm vir})]\frac{4\pi GM}{10}\rho_X(a_{\rm vir}) R_{\rm ta}^2.
\end{eqnarray}
We call using the dark energy potential $U_{X,1}$ to calculate the
total energy in the perturbation at virialisation ``fixing the energy
at turn-around'', and using $U_{X,2}$ ``fixing the energy at
virialisation''. This is demonstrated in
Fig.~\ref{fig:tot_energy_evol_w}, where we plot the total energy in an
ideal homogeneous perturbation undergoing collapse to a singularity at
$a=1$ relative to the energy at turn-around (black lines), for a
number of cosmologies. The total energy was calculated by combining
the analysis of Section~\ref{sec:dc} with the definitions of potential
energy of Eq.~\ref{eq:energy}. Obviously if the total energy were constant
we would find a horizontal line, as is the case when $w=-1$. If the
perturbation did not evolve in size, with the radius fixed at the
turn-around value, then the evolution of the total energy is shown by
the grey lines. The actual total energy within the perturbation lies
between these two extreme cases for each value of $w$ assumed.

For a more realistic perturbation that underwent virialisation rather
than collapsing to a singularity, provided that the dark energy
potential varies monotonically between $a_{\rm ta}$ and $a_{\rm vir}$,
then the change in total energy in the system between turn-around and
virialisation should still lie between these two cases: some of the
change in total energy due to the dark energy remaining homogeneous
will be converted into kinetic energy, and some will be ``lost'' to
the background.

The second scenario, using $U_{X,2}$ to ``fix'' the energy at
virialisation alters Eq.~\ref{eq:r_ratio_const_energy_ta} to give 
\be
  \left(1-[1+3w(a_{\rm vir})]\frac{q}{2y}\right)x 
    + [1+3w(a_{\rm vir})]\frac{qx^3}{y}
    =\frac{1}{2},
  \label{eq:r_ratio_const_energy_vir}	 
\ee
with $x$, $y$, \& $q$ defined as before. For $w=-1$ this equation again
reduces to the formula of \citet{lahav91}.

\begin{figure}
  \centering
  \resizebox{6.5cm}{!}{\includegraphics{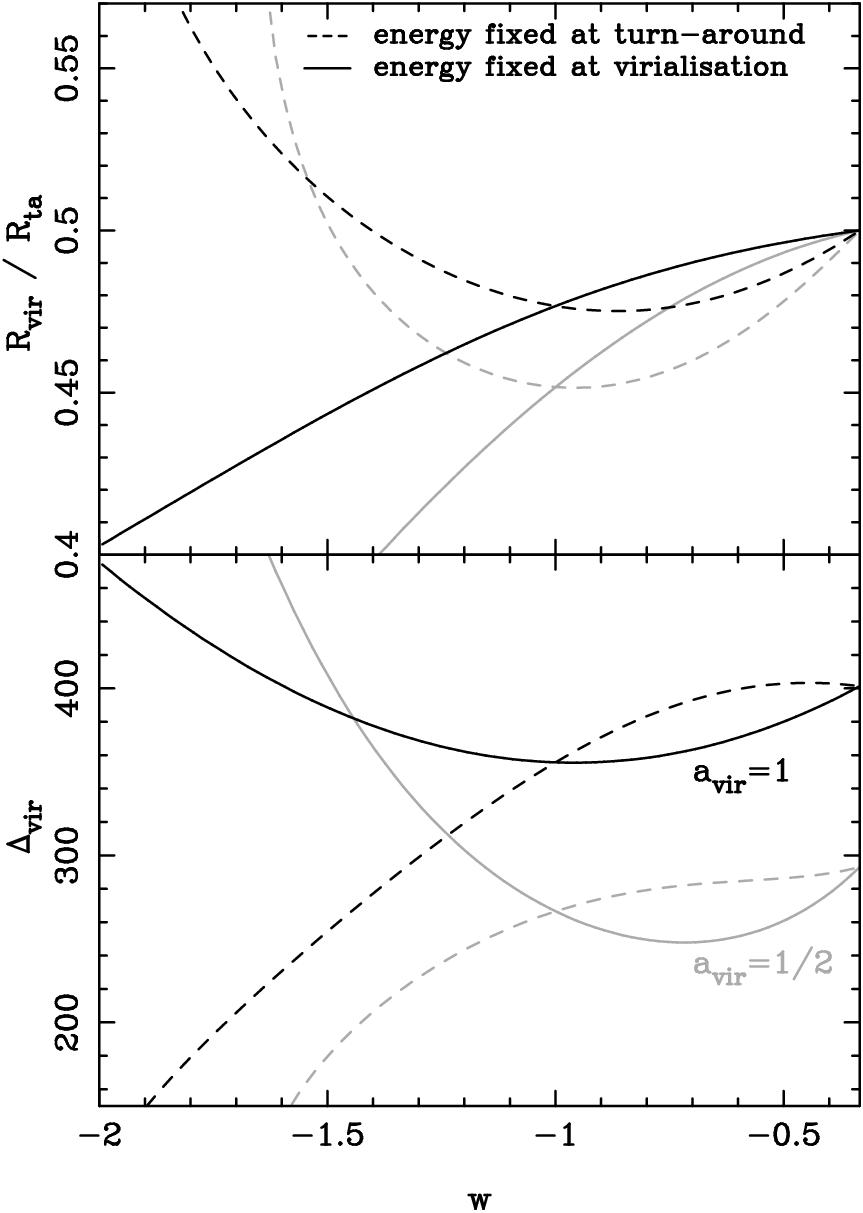}}

  \caption{Upper panel: the ratio between the virialisation radius and
   maximum turn-around radius for a spherical perturbation that
   virialises at $a=1$ (black lines) or $a=1/2$ (grey lines), assumed
   to be the same as the epoch predicted for collapse of a homogeneous
   perturbation. We assume that $\Omega_M=0.3$, $\Omega_X=0.7$, and
   plot results as a function of constant $w$. The dashed and solid
   lines show two different calculations assuming conservation of
   energy based on the dark energy density either at turn-around, or
   at virialisation. Lower panel: as upper panel, but now showing the
   overdensity of the virialised system. \label{fig:Rvir_over_Rta}}
\end{figure}

In the upper panel of Fig.~\ref{fig:Rvir_over_Rta} we plot $R_{\rm
vir}/R_{\rm ta}$ in these two cases in a cosmology with
$\Omega_M=0.3$, $\Omega_X=0.7$, as a function of constant $w$. We
assume that virialisation occurs at the collapse epoch, set to be
either the present day or $a=1/2$, for the homogeneous spherical
model. For constant $w=-1/3$ we find $R_{\rm vir}/R_{\rm ta}=1/2$ as
expected. For constant $w=-1$, the two solutions converge to the
result of \citet{lahav91}. For $-1<w<-1/3$, the component of the
energy of the perturbation in the form of dark energy changes from
turn-around to virialisation, and both energy conservation arguments
predict different values of $R_{\rm vir}/R_{\rm ta}$, although the
trend with $w$ is the same for both. However, for $w<-1$, the
solutions diverge. It is easy to see why: for $w<<-1$, the dark energy
becomes increasingly important at late times in the evolution of the
Universe. For a solution that collapses at present day (for example),
turn-around must happen at an epoch where the dark energy has yet to
be cosmologically important. It is only between turn-around and
virialisation that the dark energy becomes important. Setting the
energy to be constant based on the dark energy density at
virialisation simply continues the trend from $-1/3>w>-1$, with
$\rho_X$ increasing as $w$ decreases, leading to a decrease in the
potential due to the dark energy $U_X$, and a smaller $R_{\rm
vir}/R_{\rm ta}$ ratio. However, setting the dark energy density at
turn-around has the opposite effect: $\rho_X$ decreases, leading to
$U_X$ and $R_{\rm vir}/R_{\rm ta}$ increasing. For perturbations that
turn-around, but do not collapse, the dark energy density increases so
rapidly between turn-around and virialisation that the predicted
$R_{\rm vir}/R_{\rm ta}$ keeps increasing (although virialisation is
never reached because the perturbation can never stabilise).

In the lower panel of Fig.~\ref{fig:Rvir_over_Rta} we plot the
overdensity at virialisation $\Delta_{\rm vir}$, calculated for the
two values of the total energy. As can be seen, the result has a
very similar dependence on $w$ as described above for $R_{\rm
vir}/R_{\rm ta}$. It is clear that further analysis is required before
measurements that depend on $\Delta_{\rm vir}$ can be used to
constrain $w$. The answer may lie in numerical simulations of the time
evolution of the perturbation leading to virialisation, possibly of
the form of \citet{engineer00}, or possibly standard N-body
simulations \citep{meneghetti05,bartelmann05}. Such simulations will
need to consider the evolution of total energy with time as well as
the changing epoch of virialisation: given the possible strong effect
of the dark energy, it seems clear that the assumption that
virialisation of the perturbation occurs at the collapse time
predicted in the homogeneous case will also break down in addition to
the conservation of total energy.

\section{conclusions}

The presence of dark energy alters the way in which cosmological
structures grow, thus providing an observational signature that is
complementary to geometrical effects.  The structure growth is
dependent on the normalisation of the dark energy density $\Omega_X$,
its equation of state, and sound speed, which determines how this
component clusters. In this paper we have only considered the effect
of the first two of these properties, assuming that the dark energy
remains homogeneous on the scales of interest due to a high sound
speed. The spherical top-hat model has been used to determine the
linear growth rate and non-linear collapse overdensity threshold. The
equations provided have allowed for general $w(a)$. We have also
considered the statistics of observed structures through the mass
function and its evolution. Finally, the virialisation of
perturbations has been considered and a new argument has been
presented demonstrating the importance of the lack of energy
conservation within a perturbation. It is clear that more work is
required before observations that use arguments based on the energy in
perturbations can be used to constrain the dark energy. However, there
is clearly tremendous potential for future observations to detect the
cosmological effects of dark energy in sufficient detail to pin down
its properties.  Proving that $w\ne-1$, or indeed that $w=-1$ would be
an exciting result, and remains one of the goals of modern cosmology.

\section*{Acknowledgements}
WJP is grateful for support from PPARC.

\setlength{\bibhang}{2.0em}

\label{lastpage}

\end{document}